\documentclass[fleqn,usenatbib]{mnras}

\usepackage[T1]{fontenc}
\usepackage{ae,aecompl}

\usepackage{graphicx}	% Including figure files
\usepackage{amsmath}	% Advanced maths commands
\usepackage{amssymb}	% Extra maths symbols
\usepackage{multirow}
\usepackage{url}
\usepackage{footmisc}
\usepackage{enumitem}
\usepackage{caption}
\usepackage{hyperref}
\usepackage{mathptmx}
\usepackage{txfonts}
\usepackage{pdfpages} % include pdfs
\usepackage{pgffor} % for loops

\makeatletter
\AtBeginDocument{\let\LS@rot\@undefined}
\makeatother

\def\supplementfilename{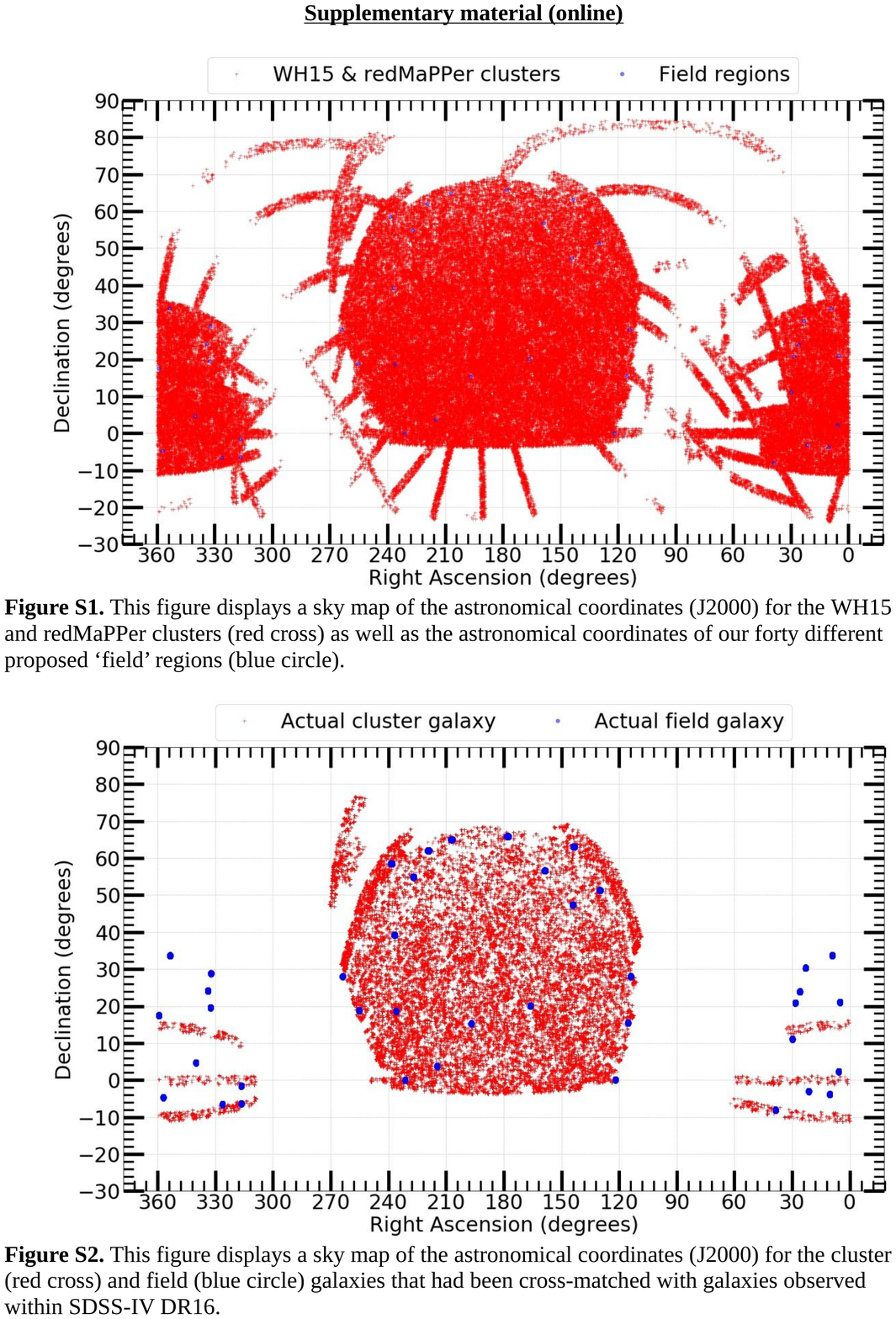}

\pdfximage{\supplementfilename}
\def\numbersupplementpages{\the\pdflastximagepages}

\newif\ifarXiv
\arXivtrue 
\title[Estimating Galaxy Cluster Richness]{AutoEnRichness: A hybrid empirical and analytical approach for estimating the richness of galaxy clusters}

\author[M. C. Chan and J. P. Stott]{
Matthew C. Chan$^{1}$ and John P. Stott$^{1}$
\\
\texttt{E-mails: m.c.chan@lancaster.ac.uk and j.p.stott@lancaster.ac.uk}
\\
$^{1}$Department of Physics, Lancaster University, Lancaster, LA1 4YB, UK
}

\date{Accepted XXX. Received YYY; in original form ZZZ}

\pubyear{2022}

\begin{document}
\label{firstpage}
\pagerange{\pageref{firstpage}--\pageref{lastpage}}
\maketitle

\begin{abstract}
\label{abstract}
We introduce AutoEnRichness, a hybrid approach that combines empirical and analytical strategies to determine the richness of galaxy clusters (in the redshift range of $0.1 \leq z \leq 0.35$) using photometry data from the Sloan Digital Sky Survey Data Release 16, where cluster richness can be used as a proxy for cluster mass. In order to reliably estimate cluster richness, it is vital that the background subtraction is as accurate as possible when distinguishing cluster and field galaxies to mitigate severe contamination. AutoEnRichness is comprised of a multi-stage machine learning algorithm that performs background subtraction of interloping field galaxies along the cluster line-of-sight and a conventional luminosity distribution fitting approach that estimates cluster richness based only on the number of galaxies within a magnitude range and search area. In this proof-of-concept study, we obtain a balanced accuracy of $83.20$ per cent when distinguishing between cluster and field galaxies as well as a median absolute percentage error of $33.50$ per cent between our estimated cluster richnesses and known cluster richnesses within $r_{200}$. In the future, we aim for AutoEnRichness to be applied on upcoming large-scale optical surveys, such as the Legacy Survey of Space and Time and $\textit{Euclid}$, to estimate the richness of a large sample of galaxy groups and clusters from across the halo mass function. This would advance our overall understanding of galaxy evolution within overdense environments as well as enable cosmological parameters to be further constrained.
\end{abstract}


%%%%%%%%%%%%%%%%%%%%%%%%%%%%%%%%%%%%%%%%%%%%%%%%%%

%%%%%%%%%%%%%%%%% BODY OF PAPER %%%%%%%%%%%%%%%%%%

\section{Introduction}
\label{introduction}

Galaxy clusters are the densest conglomerations of galaxies to have assembled in the Universe, containing tens to thousands of individual galaxies. The study of galaxy clusters is extremely important in astrophysics and cosmology research. For example, examining the mass profile of overdense environments (e.g. \citealt{mass_profile_0}; \citealt{mass_profile_1}; \citealt{mass_profile_2}; \citealt{mass_profile_3}; \citealt{mass_profile_4}; \citealt{mass_profile_5}; \citealt{mass_profile_6}; \citealt{mass_profile_7}; \citealt{mass_profile_8}); understanding the evolution of large scale structure throughout cosmic time (e.g. \citealt{cluster_evolution_0}; \citealt{cluster_evolution_1}; \citealt{cluster_evolution_2}; \citealt{cluster_evolution_3}; \citealt{cluster_evolution_4}; \citealt{cluster_evolution_5}; \citealt{cluster_evolution_6}; \citealt{cluster_evolution_7}; \citealt{cluster_evolution_8}) or mapping the distribution of clusters within the Universe (e.g. \citealt{cluster_mapping_0}; \citealt{cluster_mapping_1}; \citealt{cluster_mapping_2}; \citealt{cluster_mapping_3}; \citealt{cluster_mapping_4}; \citealt{cluster_mapping_5}; \citealt{cluster_mapping_6}; \citealt{cluster_mapping_7}; \citealt{cluster_mapping_8}). It would be beneficial for future studies if reliable, accurate and scalable methods are developed that can provide mass estimates for a large sample of clusters.
 
Historically, in order to estimate the mass of clusters, researchers have regularly turned to optical surveys for determining cluster richness, where cluster richness can provide a proxy of cluster mass such that the number of galaxies within a cluster is expected to scale with cluster mass. For example, the Abell catalogue \citep{abell_richness} was the first comprehensive large scale cluster catalogue to establish a measurement system for cluster richness, where cluster richness was defined as the number of galaxies counted within a specific radius and between two magnitude limits (i.e. the bright limit is the magnitude of the third brightest cluster galaxy whilst the faint limit is two magnitudes dimmer than the magnitude of the third brightest cluster galaxy). Similarly, the Zwicky catalogue \citep{zwicky_richness} was another comprehensive large scale cluster catalogue that established its own measurement system for cluster richness, where cluster richness was defined as the number of galaxies counted within an isopleth (i.e. the apparent boundary where the cluster density is twice that of the field density) and also between two magnitude limits (i.e. the bright end limit is the magnitude of the brightest cluster galaxy whilst the faint end limit is three magnitudes dimmer than the magnitude of the brightest cluster galaxy). We note that our definition of richness in this paper is the number of cluster galaxies up to an absolute magnitude faint-end $r$ filter limit of $-20.5$ and within an $r_{200}$ radius.

In more recent times, a variety of automated methods have been developed that enable cluster mass or richness to be estimated without the need for extensive manual processing, such as utilising linking algorithms within redshift space (e.g. \citealt{group_selection_algorithm_0}; \citealt{group_selection_algorithm_1}; \citealt{group_selection_algorithm_2}; \citealt{group_selection_algorithm_3}; \citealt{whl12}; \citealt{group_selection_algorithm_4}; \citealt{group_selection_algorithm_5}), employing template fitting algorithms within colour-magnitude space (e.g. \citealt{matched_filter_algorithm_0}; \citealt{matched_filter_algorithm_1}; \citealt{matched_filter_algorithm_2}; \citealt{matched_filter_algorithm_3}; \citealt{photometric_cluster_members_0}; \citealt{redmapper}) or training machine learning algorithms on observational/simulated measurements to indirectly estimate cluster mass (e.g. \citealt{cluster_mass_machine_learning_0}; \citealt{cluster_mass_machine_learning_1}; \citealt{cluster_mass_machine_learning_2}; \citealt{cluster_mass_machine_learning_3}; \citealt{cluster_mass_machine_learning_4}; \citealt{cluster_mass_machine_learning_5}; \citealt{cluster_mass_machine_learning_6}).

Alternative approaches to determine cluster mass commonly include X-ray, caustic and weak lensing methods. From which, X-ray methods assume that the intracluster gas within a cluster is under hydrostatic equilibrium in order to calculate the cluster mass required to produce the observed X-ray emissions, based on X-ray temperature and surface brightness measurements (e.g. \citealt{xray_0}; \citealt{xray_1}; \citealt{xray_2}); caustic methods assume a cluster has spherical symmetry in order to calculate the cluster mass required to generate an estimated average escape velocity for cluster galaxies, based on galaxy position and velocity measurements (e.g. \citealt{caustics_0}; \citealt{caustics_1}; \citealt{caustics_2}); whilst weak lensing methods make no physical assumptions about a cluster to estimate the cluster mass required to produce the observed gravitational lensing of light from background objects, based on light distortion and magnification measurements (e.g. \citealt{weak_lensing_0}; \citealt{weak_lensing_1}; \citealt{weak_lensing_2}). Although, these methods have somewhat time-consuming and expensive prerequisites (e.g. conducting deep X-ray observations, requiring complete spectroscopic analysis, obtaining high quality image data for performing weak lensing analysis), whereas methods involving optical photometry are typically quicker and cheaper to obtain and analyse the resultant data.

We note that determining cluster richness from the direct counting galaxies within a cluster is limited by the projection effect (\citealt{projection_effect_0}; \citealt{projection_effect_1}; \citealt{projection_effect_2}; \citealt{projection_effect_3}; \citealt{projection_effect_4}), where \cite{abell_catalog_completeness} estimated that approximately one third of the clusters in the Abell catalogue may have had their richnesses severely misestimated due to contamination from the projection effect. This effect arises when foreground or background galaxies are in the same line-of-sight as the cluster itself, which means it is difficult to accurately associate galaxies to a cluster unless spectroscopic redshifts for each galaxy are known. However, this is time-consuming especially when working with large sample sizes, as it is dependent on the preciseness of the distance measurement required. 

In the literature, various statistical and non-statistical background subtraction methods have been utilised to address the projection effect when obtaining counts of cluster galaxies without the need for distance measurements. One typical way is to count the number of field galaxies within a known control field sample, which can be used as a direct reference to subtract a proportional number of galaxies from a cluster's overall population to account for field galaxies (e.g. \citealt{control_field_0}; \citealt{control_field_1}; \citealt{control_field_2}). Another way is to define an annuli around the apparent outer perimeter of a cluster, which assumes that the annuli is far enough away to likely not contain cluster galaxies, such that a proportional number of galaxies can be subtracted from a cluster's overall population to account for field galaxies (e.g. \citealt{schechter_alpha_5}; \citealt{annuli_0}; \citealt{annuli_1}). A further approach is to establish colour cuts for differentiating between cluster and field galaxies, where most of the galaxies within a cluster will appear to have similar colours especially if they are at the same redshift (i.e. red-sequence galaxies), whilst galaxies in the field will appear more randomised in terms of colour, especially if they are at different redshifts (e.g. \citealt{colour_cut_0}; \citealt{colour_cut_1}; \citealt{colour_cut_2}). However, the approaches described here may not provide a robust or precise enough background subtraction, which is essential for accurately estimating cluster richnesses, due to these methods either being statistical or not assessing the true membership status of each cluster galaxy.

For this paper, we describe in detail a novel hybrid method, nominally known as AutoEnRichness, to perform background subtraction and estimate cluster richnesses by employing a multi-stage machine learning algorithm and a conventional luminosity distribution fitting approach respectively. The first key stage of our hybrid method involves training the multi-stage machine learning algorithm to differentiate between cluster and field galaxies. This approach is completely data-driven to automatically capture underlying relationships for maximising the accuracy of cluster galaxy identification. The second key stage of our hybrid method involves learning the best fit parameters for a luminosity distribution fitting function to enable the estimation of cluster richness from the luminosity distribution of individual clusters. This approach has a strong theoretical basis that depends only on the brightness of the cluster galaxy population within a given search radius of a cluster. Our proposed strategy will be beneficial to provide researchers in the field with well-founded estimates of cluster richness as well as consistency and robustness against line-of-sight effects to mitigate severe contamination. 

We present this paper with the following structure. Firstly, in \S\ref{sec:methodology} we divide our methodology into five subsections, where \S\S\ref{sec:prepare_photometry_for_background_subtraction} describes the preparation of a photometric dataset to train a background subtraction model; \S\S\ref{sec:autoencoder_algorithm} describes the mechanisms of a multi-stage machine learning algorithm that is used as our background subtraction model; \S\S\ref{sec:scaling_relation_r200} describes our strategy for establishing a scaling relation to estimate $r_{200}$ of clusters; \S\S\ref{sec:prepare_photometry_to_predict_richness} describes the preparation of a photometric dataset to train a luminosity distribution fitting function and \S\S\ref{sec:luminosity_distribution_fitting} describes the mechanisms of a luminosity distribution fitting function to estimate cluster richness. In \S\ref{sec:results} we outline our results across three subsections, where \S\S\ref{sec:model_tuning_analyses} describes the model tuning analyses of our learned background subtraction model, scaling relation and luminosity distribution fitting function; \S\S\ref{sec:analyses_with_test_sets} describes the overall performance of our methodology on unseen clusters in various test sets and \S\S\ref{sec:examining_feature_importance} describes the importance of input features to our background subtraction model. Lastly, \S\ref{sec:discussion} discusses our findings and \S\ref{sec:conclusion} summarises this paper.

We assumed the following $\Lambda$CDM cosmological parameters $H_{0} = 71 \ \text{km} \ \text{s}^{-1} \ \text{Mpc}^{-1}$, $\Omega_{m} = 0.27$ and $\Omega_{\Lambda} = 0.73$.

\section{Methodology}
\label{sec:methodology}

\begin{figure}
\centering
	\includegraphics[width=\linewidth] {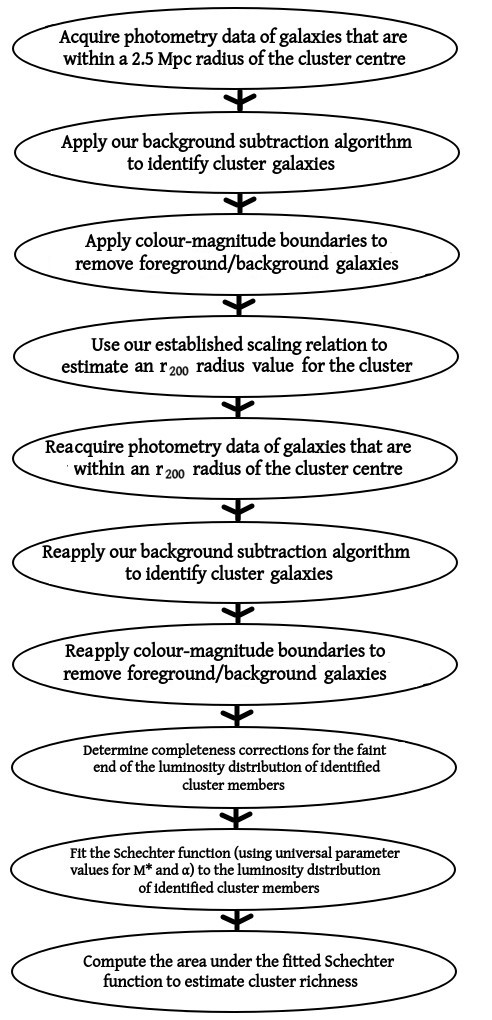}
    \caption{This figure shows a flowchart of the various steps in our multi-stage method to estimate the richness of a cluster, where the start of the flowchart is the first step whilst the end of the flowchart is the last step.}
    \label{fig:method_flowchart}
\end{figure}

A brief outline of our multi-stage method to estimate the richness of a cluster can be seen in Figure \ref{fig:method_flowchart}. From which, the following subsections will describe our workflow in more detail.

\subsection{Preparation of a photometric dataset to train a background subtraction model}
\label{sec:prepare_photometry_for_background_subtraction} 

To train our background subtraction model, we employed cluster galaxies that were identified by the \cite{photometric_cluster_members_0} cluster catalogue (hereafter we refer to this cluster catalogue as the AMF11 catalogue) with an estimated photometric redshift between $0.1 \leq z \leq 0.35$. We note that the AMF11 catalogue applied matched filters\footnote[1]{The matched filters were constructed from modeling positional, brightness and redshift information of cluster and field galaxy distributions.} to galaxies observed in the Sloan Digital Sky Survey Data Release 6 (SDSS-II DR6, \citealt{sdss_dr6}), where clusters were detected from maximising the likelihood of the matched filters whilst cluster galaxy membership identification was based on the proximity of the galaxy from the cluster center within $r_{200}$\footnote[2]{We refer to $r_{200}$ as the radius containing a mean density that is two hundred times greater than the critical density of the Universe.} as well as whether the likelihood difference (i.e. the difference in likelihood of detecting a cluster with and without the presence of the galaxy) was above a specified threshold. The reason we decided to use the cluster galaxies from the AMF11 catalogue was because they assessed the cluster membership status of each galaxy based on their contribution to a combination of various cluster profiles (i.e. radial surface density, luminosity and redshift). In addition, their selection method does not discriminate between `blue' and `red' cluster galaxies, which means it is representative of different galaxy types in clusters.

We cross-matched these cluster galaxies with galaxies observed in the Sloan Digital Sky Survey Data Release 16 (SDSS-IV DR16, \citealt{sdss_dr16}) to obtain the following fifteen features
that are based on SDSS-IV DR16 photometry\footnote[3]{We employed full-sky dust reddening maps (\citealt{dust_extinction_0}; \citealt{dust_extinction_1}) to account for galactic extinction.}: $u$, $g$, $r$, $i$, $z$, $u-g$, $g-r$, $r-i$, $i-z$, $u-r$, $g-i$, $r-z$, $u-i$, $g-z$ and $u-z$. For a cluster galaxy to be successfully cross-matched, the input astronomical coordinates must be within $1$ arcsecond from the astronomical coordinates of a galaxy within SDSS-IV DR16 as well as satisfying additional observing flags. These flags are as follows: the observed object should be a `primary' observation\footnote[4]{We note that SDSS uses the term `primary' to refer to the best imaging observation recorded for a survey object if it was seen multiple times during an observing run in an SDSS plate, whilst other observations of the object are known as `secondary'.} and must be classified as a galaxy object type by the SDSS photometric pipeline. We also ensured that our cluster galaxy sample only contained galaxies with a unique SDSS object identifier to prevent accidentally including galaxies that may have been selected multiple times within our search radius due to very small angular separation between overlapping line-of-sight galaxies and errors in the astrometry. In addition, we did not include cluster galaxies that were within $1646$ arcseconds (i.e. $3$ Mpc at $z = 0.1$) of a subsample (see \S\S\ref{sec:scaling_relation_r200} for further details) of cross-matched\footnote[5]{This involved identifying clusters that were within $70$ arcseconds (i.e. $250$ kpc at $z = 0.225$) of each other in astronomical coordinate space and within $\pm 0.04(1 + z)$ (see \cite{photometric_redshift_gap} for further explanation) of each other in redshift space. In addition, the clusters had to be observed within SDSS-IV DR16 between a redshift range of $0.1 \leq z \leq 0.35$.} clusters from the \cite{wh15} cluster catalogue (hereafter we refer to clusters from this catalogue as WH15 clusters) and \cite{redmapper} cluster catalogue (hereafter we refer to clusters from this catalogue as redMaPPer clusters). This ensured that these clusters remained unseen for later usage in \S\S\ref{sec:scaling_relation_r200}. Furthermore, we applied a cut within colour-magnitude space (i.e. if greater than the $99.75th$ percentile in $r$ and $g-r$) to remove any cluster galaxies that still appeared to have spurious photometry. It should be noted that throughout this work, we used $r$ and $g-r$ to visualise cluster and field galaxies in colour-magnitude diagrams due to $g-r$ straddling the $4000\AA$ break of cluster galaxies in our working redshift range.

Correspondingly, we also required a field galaxy\footnote[6]{We refer to interloping galaxies along a clusters line-of-sight as field galaxies.} sample to train our background subtraction model to differentiate between cluster and field galaxies. However, we were unable to find a sizable catalogue containing identified field galaxies. This meant that we had to manually search for `field' regions that did not visually appear to contain clusters from the full WH15 and redMaPPer cluster catalogues. This resulted in the identification of forty different `field' regions, where the resultant astronomical sky map displaying the position of clusters and our proposed `field' regions can be seen in Figure S1 (available online). We sampled galaxies from SDSS-IV DR16 that were within these `field' regions. This involved applying a $1372$ arcseconds (i.e. $2.5$ Mpc at $z = 0.1$) search radius on each of the `field' regions as well as reusing the same observing flags mentioned earlier within this section to obtain our field galaxy sample. The astronomical coordinates and number of observed field galaxies for each `field' region are provided in Table \ref{tab:field_coords}. We did not include field galaxies that were within $10$ arcseconds from the cluster galaxies in the AMF11 catalogue to remove cluster galaxies that may have accidentally been included as part of the field regions. We also did not include field galaxies that were within $1646$ arcseconds (i.e. $3$ Mpc at $z = 0.1$) of the same subsample of cross-matched WH15 and redMaPPer clusters mentioned earlier within this section to ensure that the clusters remained unseen for later usage in \S\S\ref{sec:scaling_relation_r200}. Furthermore, we removed any field galaxies that were not within the same region of colour-magnitude space as our cluster galaxy sample, based on the observed minimum and maximum values for the cluster galaxies in $r$ and $g-r$. This was intended to encourage our background subtraction model to learn to be more proficient at classifying galaxies with similar photometric properties. Subsequently, this yielded a total of $83315$ field galaxies that had the same fifteen photometry features as our cluster galaxy sample. For this paper, we assumed that these field galaxies can be considered as `actual' field galaxies.

\begin{table}
\centering
	\begin{tabular}{|c|c|c|}
		\hline
	    \textbf{Right ascension} & \textbf{Declination} & \textbf{Number of observed galaxies} \\
	    \textbf{(degrees)} & \textbf{(degrees)} & \\
		\hline
5.10408	& 21.0611 & 4488 \\
5.77875	& 2.30955 & 7049 \\
9.14699	& 33.7121 & 4030 \\
10.4395	& -3.84934 & 5377 \\
21.3533	& -3.02485 & 5595 \\
23.0452	& 30.3202 & 5664 \\
26.0012	& 23.8818 & 5024 \\
28.3683	& 20.8955 & 2748 \\
29.8579	& 11.0895 & 5108 \\
38.6421	& -8.10943 & 5168 \\
113.919	& 28.0332 & 5223 \\
115.323	& 15.458 & 6310 \\
121.916	& 0.109439 & 5518 \\
129.971	& 51.2851 & 4753 \\
143.518	& 63.0839 & 4564 \\
144.097	& 47.3992 & 4929 \\
158.633	& 56.6154 & 5982 \\
166.196	& 20.0867 & 5861 \\
177.951	& 65.8863 & 5733 \\
196.86 & 15.2355 & 5885 \\
207.127	& 65.0462 & 5711 \\
214.632	& 3.77235 & 6109 \\
219.239	& 62.0499 & 6009 \\
226.958	& 54.9023 & 5346 \\
231.26 & -0.0179 & 6696 \\
235.973	& 18.662 & 7669 \\
236.823	& 39.221 & 6421 \\
238.422	& 58.5216 & 6760 \\
255.233	& 18.8401 & 6079 \\
263.834	& 28.0521 & 3565 \\
316.524	& -6.36219 & 6273 \\
316.54 & -1.56952 & 5201 \\
326.258	& -6.56724 & 6597 \\
332.263	& 28.8328 & 4580 \\
332.498	& 19.5978 & 4043 \\
333.873	& 24.14	& 4214 \\
340.098	& 4.71022 & 4252 \\
353.561	& 33.6502 & 4239 \\
357.0789 & -4.69765	& 5103 \\
359.369	& 17.4744 & 5809 \\
		\hline
	\end{tabular}
	\caption{This table contains the astronomical coordinates (J2000) and number of observed galaxies that were sampled from our forty different proposed `field' regions using a $1372$ arcseconds (i.e. $2.5$ Mpc at $z = 0.1$) search radius. We note that the number of observed galaxies does not include field galaxies that were within $10$ arcseconds of the galaxies in our cluster galaxy sample nor did we include field galaxies that were within $1646$ arcseconds (i.e. $3$ Mpc at $z = 0.1$) of a subsample of cross-matched WHL and redMaPPer clusters.}
	\label{tab:field_coords}
\end{table}            

We decided to set the redshift values of galaxies in our cluster and field galaxy samples to be based only on the photometric redshifts estimated by SDSS-IV DR16. This would enable a more straightforward comparison between the redshift distributions of both samples if they were measured via the same approach. We note that SDSS-IV DR16 applied the kd-tree nearest neighbor fit algorithm (see \cite{sdss_photometric_redshifts} for further details) to estimate the photometric redshifts of individual galaxies. We also used their estimated photometric redshifts to further constrain galaxies within our cluster galaxy sample to only be between a redshift range of $0.1 \leq z \leq 0.35$, whereas galaxies within our field galaxy sample were not redshift restricted to mimic field galaxies appearing along the line-of-sight of clusters. Although, we note that galaxies were not required to have photometric redshifts available to be included in our field galaxy sample. In addition, we computed the $r$ filter absolute magnitudes for the cluster and field galaxies based on their photometric redshifts and corresponding K corrections\footnote[7]{In order to estimate the amount of K correction required, we performed linear interpolation between redshift and $r$ filter K corrected values determined from a simple stellar population model (see \cite{k_correction} for further details).}.

\begin{figure}
\centering
	\includegraphics[width=\linewidth] {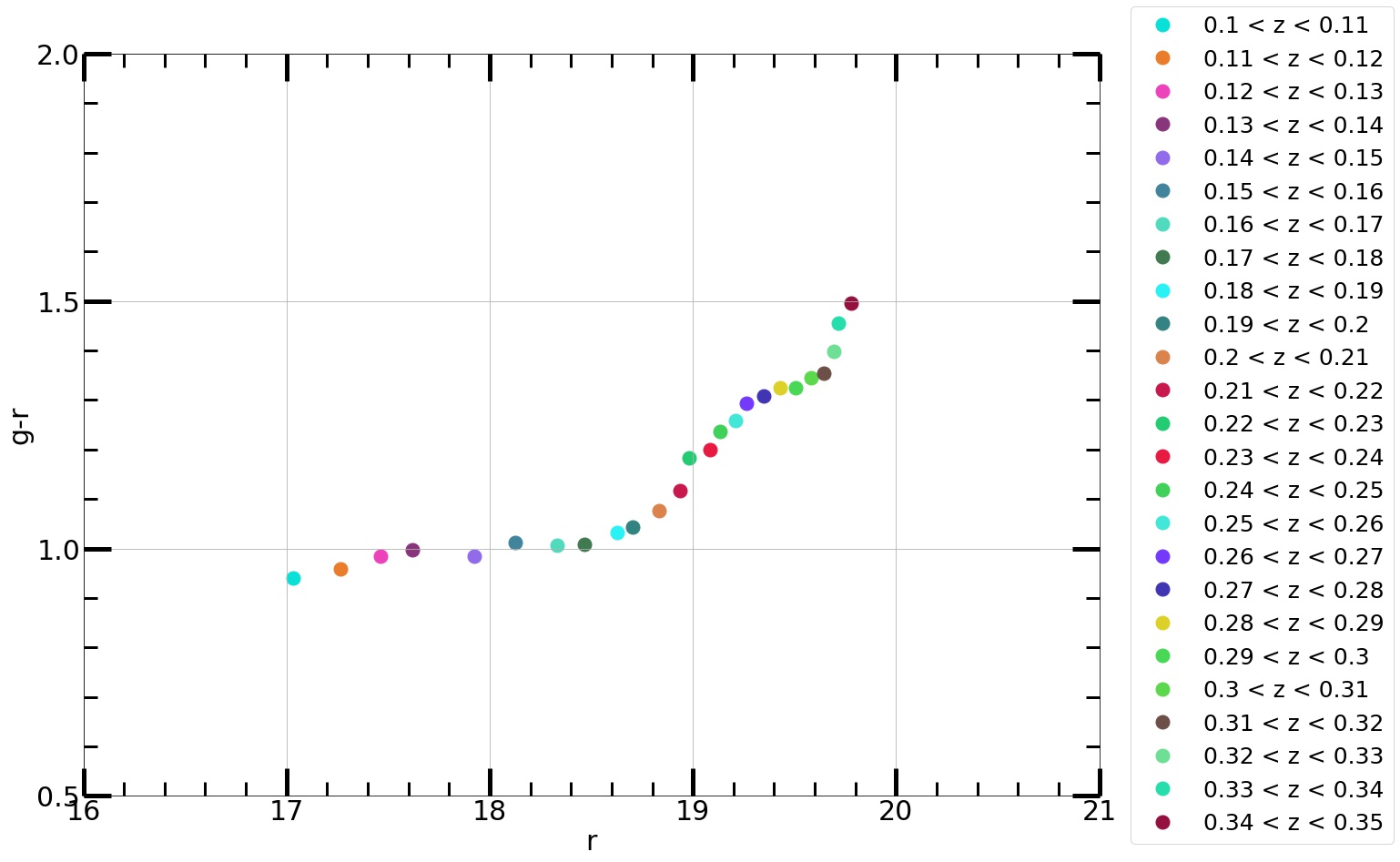}
    \caption{This figure shows a colour-magnitude diagram (using apparent magnitudes) of the median $r$ and $g-r$ for cluster galaxies at different redshift intervals from our cluster galaxy sample.}
    \label{fig:cmd_redshift_relation}
\end{figure}

\begin{figure}
\centering
	\includegraphics[width=\linewidth] {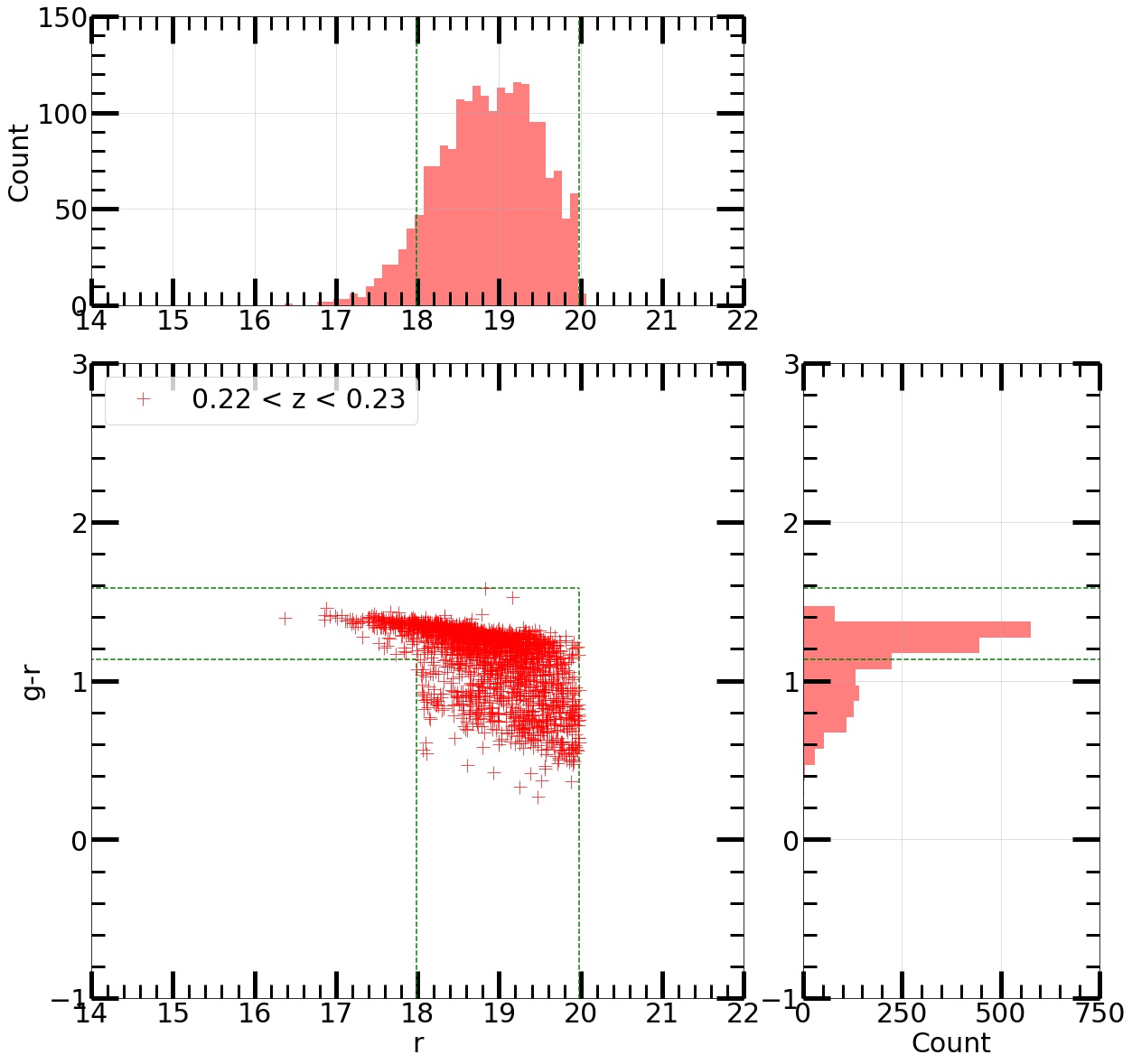}
    \caption{This figure shows an example of the colour-magnitude boundaries (green dotted lines) for cluster galaxies (red cross) between $0.22 < z < 0.23$ from our cluster galaxy sample, where only galaxies that are between the colour-magnitude boundaries will be considered as part of a cluster at that redshift.}
    \label{fig:cmd_boundaries_example}
\end{figure}

We note that our background subtraction model will learn to identify all cluster galaxies between a redshift range of $0.1 \leq z \leq 0.35$, which may result in overcounting of cluster galaxies within a cluster if there are other clusters along the line-of-sight. To limit this effect, we decided to establish colour-magnitude boundaries within colour-magnitude space when applying our background subtraction model. These boundaries are designed to capture the majority of the population of cluster galaxies at specific redshifts. We first computed the median values of $r$ and $g-r$ for cluster galaxies in our cluster galaxy sample across redshift intervals of $\pm 0.005$ that are centered in redshift bins from $0.105$ to $0.345$ with step sizes of $0.01$, as shown in Figure \ref{fig:cmd_redshift_relation}. We then manually determined appropriate lower and upper boundaries of $ r_{median} \ - \ 0.01 \leq r_{median} \leq r \ +\ 0.4$ and $g-r_{median} \ - \ 0.05 \leq g-r \leq g-r_{median} \ + \ 0.4$ for each redshift bin. This would result in `L-shaped' boundaries around the cluster galaxies at a given redshift, where an example of the `L-shaped' boundaries for cluster galaxies at $z = 0.225$ is shown in Figure \ref{fig:cmd_boundaries_example}. We then applied these colour-magnitude boundaries to our cluster galaxy sample across redshift bin sizes of $0.01$ to remove any cluster galaxies that were not within the colour-magnitude boundaries at their respective redshift. Subsequently, this yielded a total of $60663$ cluster galaxies that were available to train our background subtraction model. For this paper, we assumed that these cluster galaxies can be considered as `actual' cluster galaxies.

\begin{figure}
\centering
	\includegraphics[width=\linewidth] {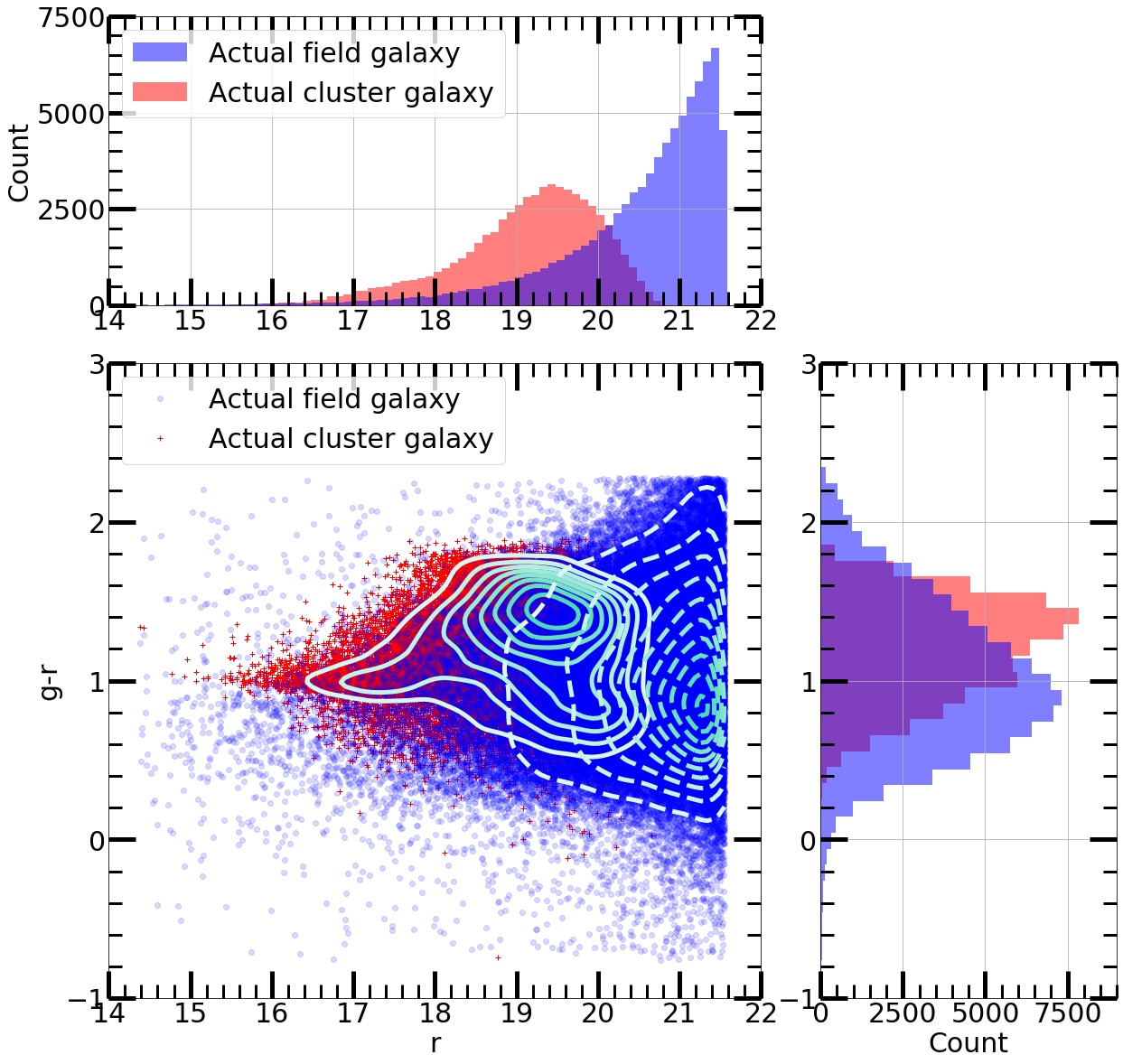}
    \caption{This figure shows colour-magnitude diagrams (using apparent magnitudes) of the cluster (red cross) and field (blue circle) galaxies in our cluster and field galaxy samples that were observed within SDSS-IV DR16. The non-dashed contour lines represent the density of data points for cluster galaxies whilst the dashed contour lines represent the density of data points for field galaxies.}
    \label{fig:cmd_cluster_vs_field}
\end{figure}

\begin{figure}
\centering
	\includegraphics[width=0.92\linewidth]{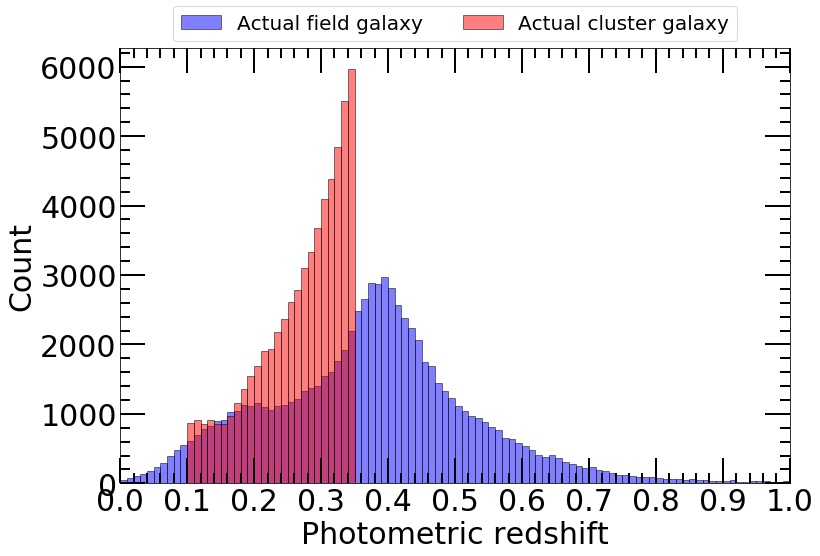}
	\includegraphics[width=0.92\linewidth]{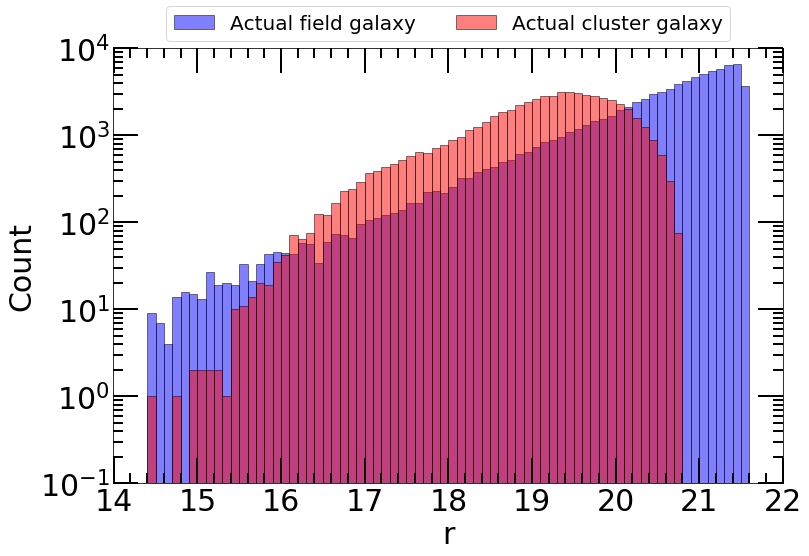}
	\includegraphics[width=0.92\linewidth]{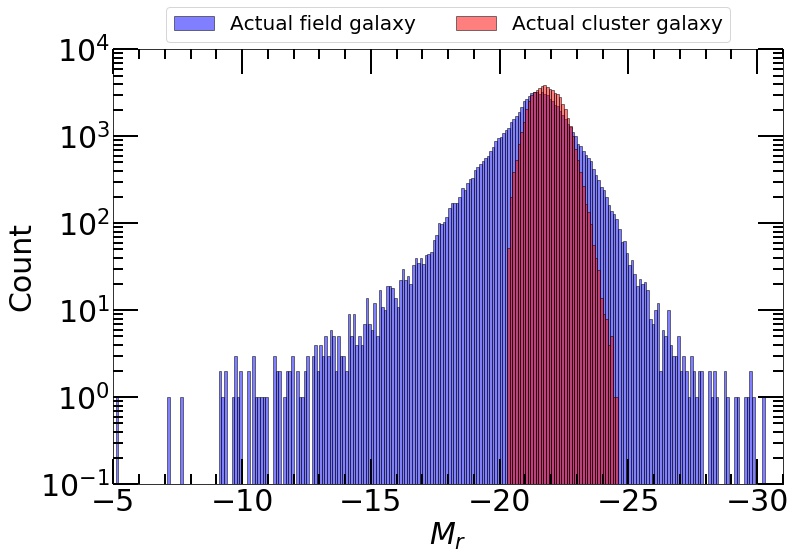}
    \caption{This figure shows histograms of the photometric redshift (top image), $r$ filter apparent magnitude (middle image) and $r$ filter absolute magnitude (bottom image) of galaxies in our cluster (red) and field (blue) galaxy samples after being cross-matched with galaxies observed within SDSS-IV DR16, where the cluster galaxies had to be between a redshift range of $0.1 \leq z \leq 0.35$. It should be noted that we only display field galaxies that had an available photometric redshift in the top and bottom images.}
    \label{fig:photometric_cluster_and_field_galaxies_histogram}
\end{figure}

In Figure S2 (available online), it can be seen that our cluster and field galaxies were taken from different areas across SDSS-IV DR16. This meant that our cluster and field galaxy samples were likely to be representative of the whole population of cluster (between a redshift range of $0.1 \leq z \leq 0.35$) and field galaxies. Moreover, in Figures \ref{fig:cmd_cluster_vs_field}, S3 and S4 (available online), it can be seen that our field galaxy sample had an overall noticeable disparity to our cluster galaxy sample within colour-magnitude space. This somewhat validated our approach for obtaining the field galaxies given the underlying differences in photometry between the majority of the cluster and field galaxies. Although, we also observed some overlap of the `blue' and faint cluster galaxies with bright field galaxies. We expect that it may be more difficult for our background subtraction model to differentiate between the galaxy classes within these overlap regions of colour-magnitude space. Furthermore, in Figure \ref{fig:photometric_cluster_and_field_galaxies_histogram} we display the photometric redshift, $r$ filter apparent magnitude and $r$ filter absolute magnitude distributions of galaxies in our cluster and field galaxy samples. It can be seen that we had fewer cluster and field galaxies at lower redshifts when compared to those at higher redshifts. This indicated that we would need to sample equally across different redshifts to prevent our background subtraction model from being biased towards any particular redshift. We note that the number of field galaxies decreased significantly after $z = 0.4$ due to the observing limitations of SDSS-IV DR16 at higher redshifts, where SDSS-IV DR16 had a $r$ filter limiting magnitude of $22.2$. We also noticed that there was a gradual drop in the number of cluster and field galaxies towards fainter magnitudes due to the incompleteness of cluster galaxies in the AMF11 catalogue and observing limitations of SDSS-IV DR16 respectively.

\begin{figure}
\centering
	\includegraphics[width=0.92\linewidth] {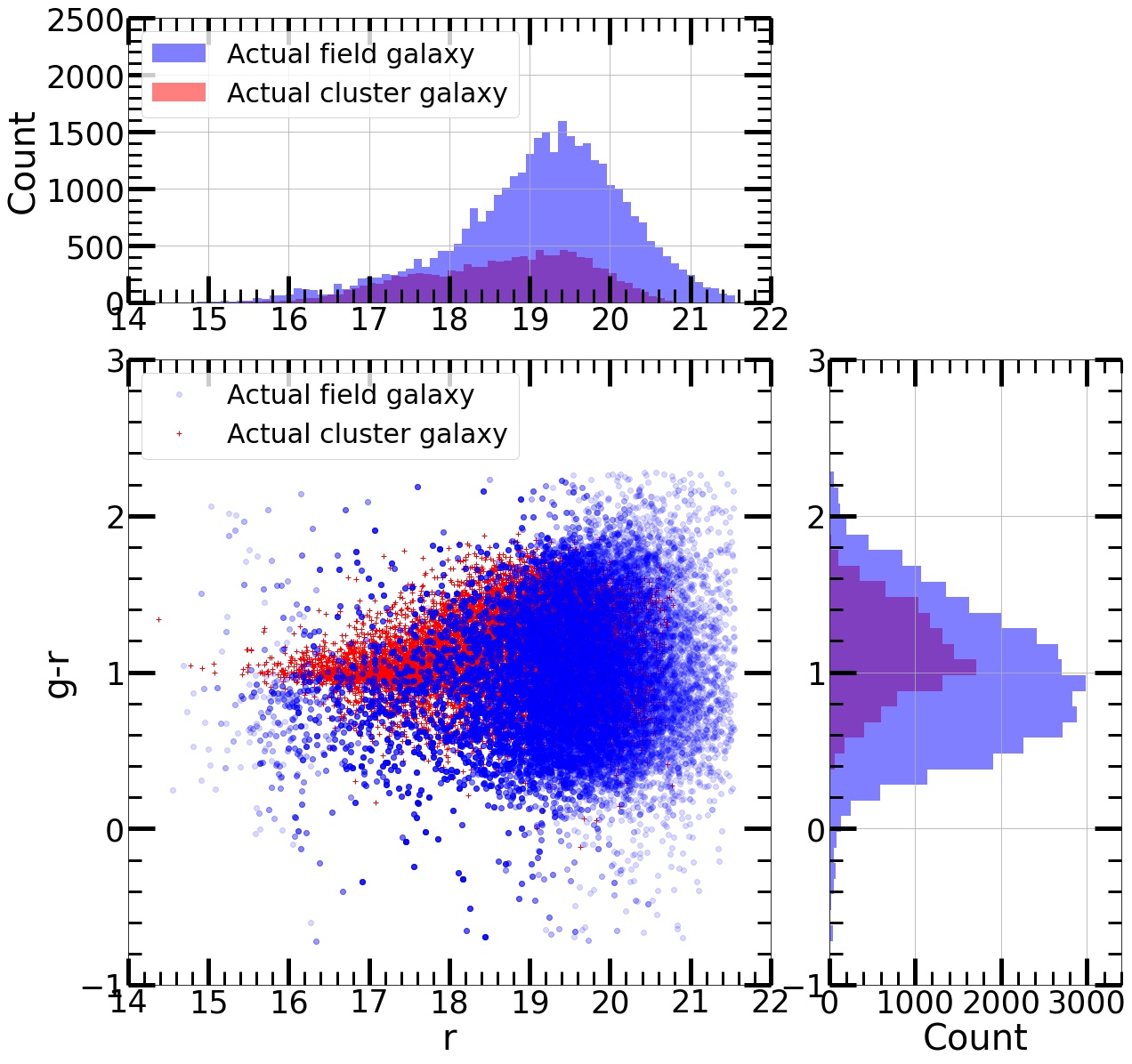}
	\includegraphics[width=0.92\linewidth] {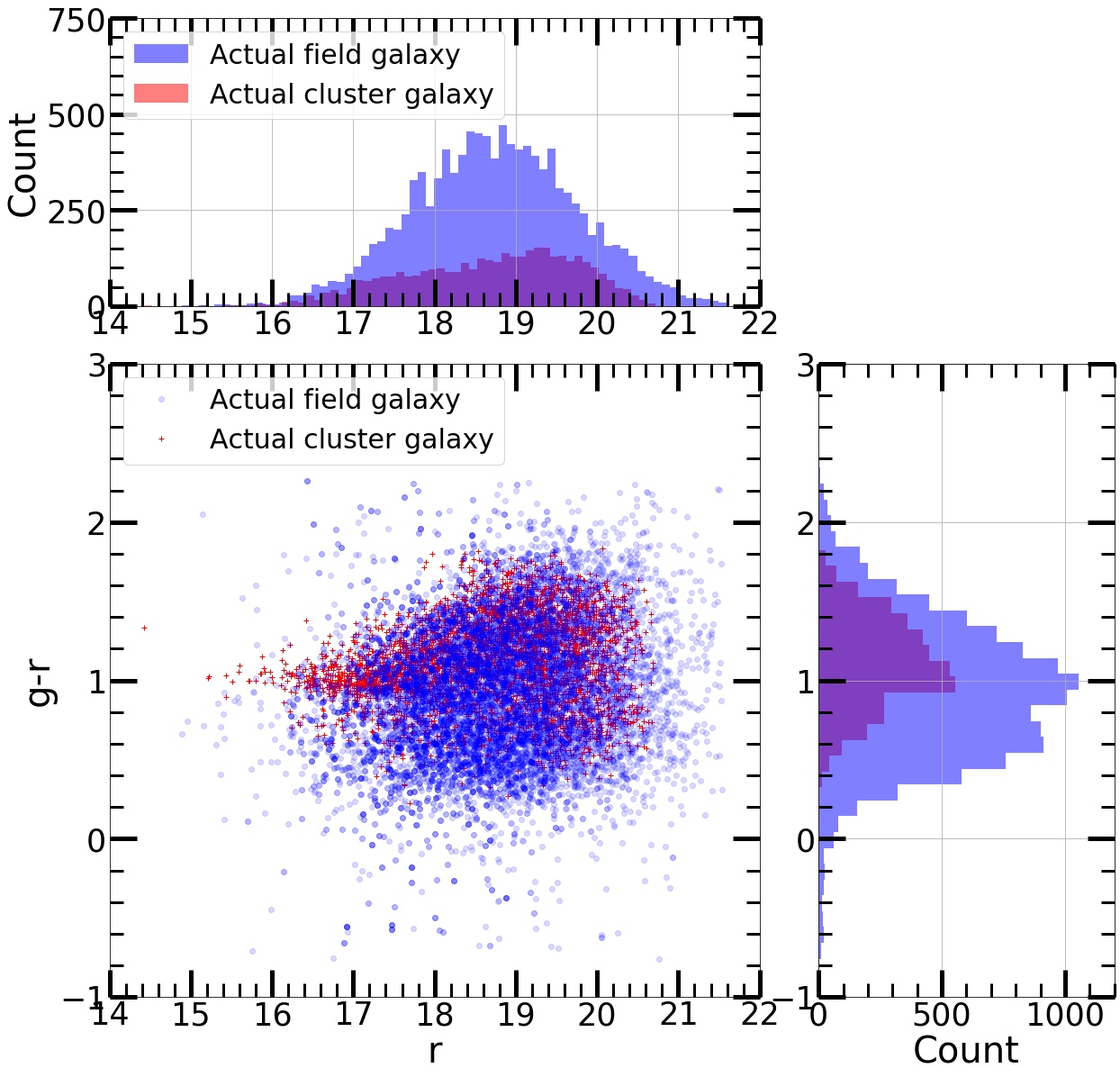}
	\includegraphics[width=0.92\linewidth] {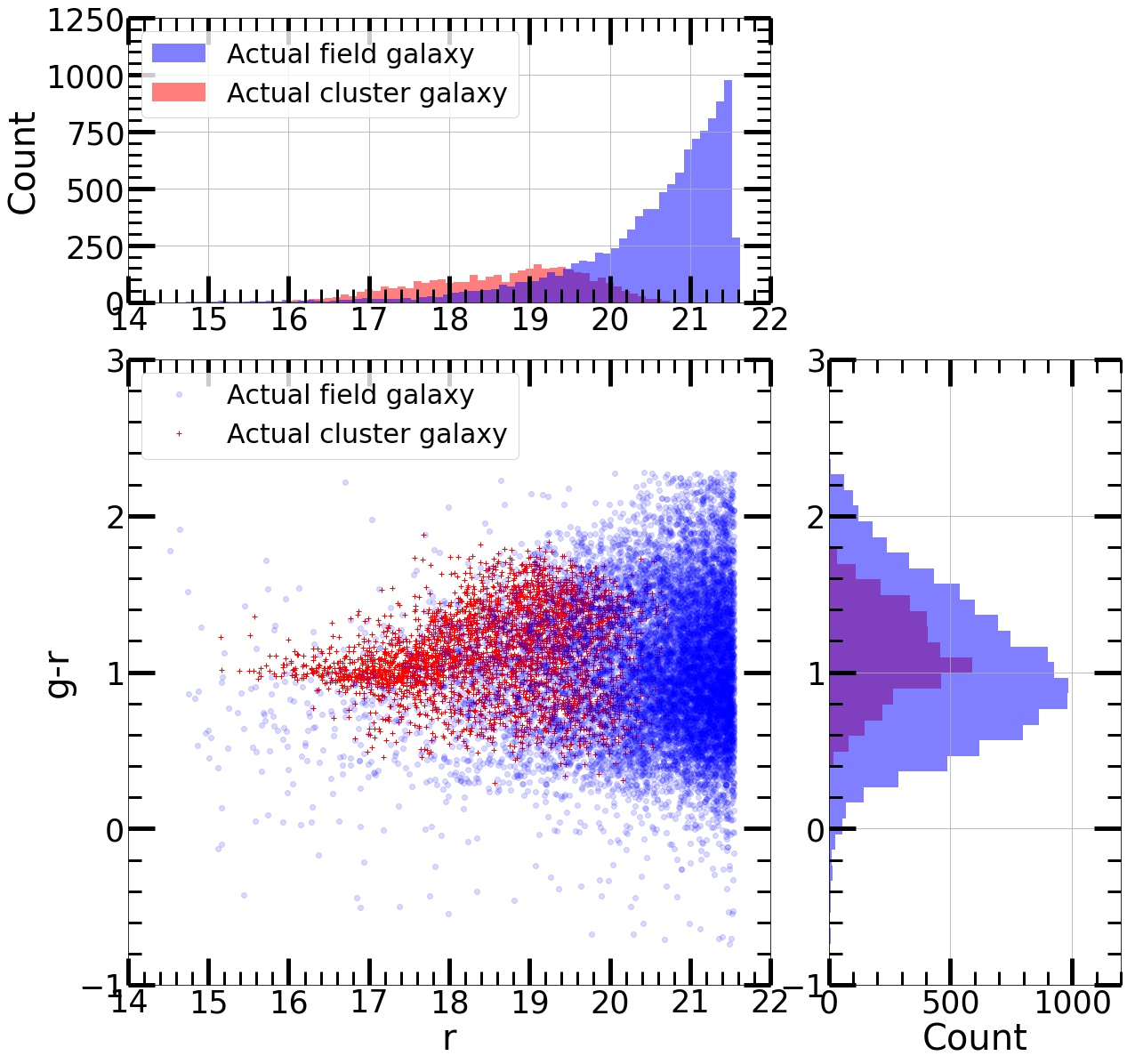}
    \caption{This figure shows colour-magnitude diagrams (using apparent magnitudes) of the cluster (red cross) and field (blue circle) galaxies in our training (top image), validation (middle image) and test (bottom image) sets that were observed within SDSS-IV DR16.}
    \label{fig:cmd_cluster_vs_field_train_val_test}
\end{figure}

Finally, we partitioned our cluster and field galaxy samples into three different subsets, known as the training, validation and test sets. In particular, the training set would be used to train our background subtraction model, the validation set would be used to tune its hyper-parameters and the test set would be used to obtain an unbiased estimate of the predictive performance of our background subtraction model. This involved randomly selecting $450$, $150$ and $150$ cluster galaxies within fixed redshift bin sizes of $0.01$ across a redshift range of $0.1 \leq z \leq 0.35$ to be within our training, validation and test sets, which resulted in a total of $11250$, $3750$ and $3750$ cluster galaxies respectively. We also randomly selected $33750$, $11250$ and $11250$ field galaxies to be within our training, validation and test sets respectively. It should be noted that we applied sampling weights\footnote[8]{The amount of sampling weightage applied to each field galaxy in our training and validation sets was based on the resultant likelihood of the $r$ filter apparent magnitude for the field galaxy under a normal distribution that was constructed from the mean and standard deviation of the $r$ filter apparent magnitudes of cluster galaxies in our training and validation sets. We also shifted the computed means by $-1$ in our training and validation sets to ensure that the cluster and field galaxy distributions overlapped at all $r$ filter apparent magnitudes. In addition, we note that sampling with replacement was used when selecting field galaxies to be within our training and validation sets.} when selecting field galaxies to be within our training and validation sets to ensure that the $r$ filter apparent magnitudes of the field galaxies overlapped with the $r$ filter apparent magnitudes of the cluster galaxies. This would expose our background subtraction model to a larger proportion of the more difficult instances (i.e. cluster and field galaxies that had very similar photometry) during its training. Furthermore, we wanted our training, validation and test sets to remain as realistic as possible. As such, we permitted the number of field galaxies to outnumber (i.e. we assumed that having three field galaxies for every cluster galaxy was appropriate) the number of cluster galaxies with these sets. Although, we only permitted random sampling (i.e. equal sampling weightage) of field galaxies in our test set. These properties can be seen in Figure \ref{fig:cmd_cluster_vs_field_train_val_test}. 

\subsection{Using a multi-stage machine learning algorithm to perform background subtraction}
\label{sec:autoencoder_algorithm}

\begin{figure}
\centering
	\includegraphics[width=\linewidth]{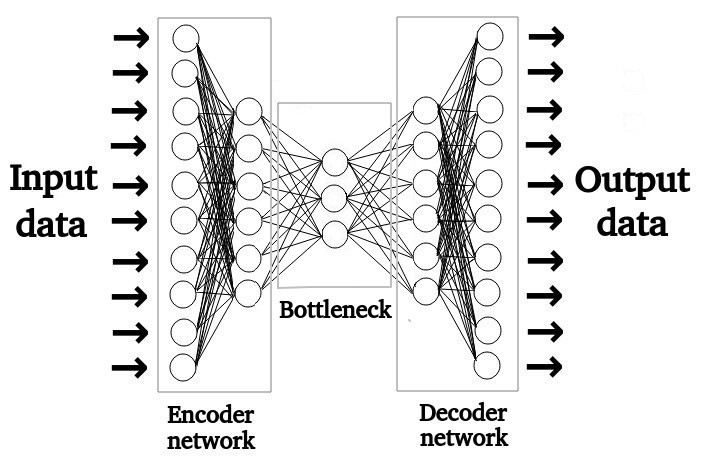}
    \caption{This figure shows an example of the architecture layout for a typical AE. The AE is composed of three main stages that are known as the encoder network, bottleneck and decoder network, where the nodes in each hidden layer are fully-connected to the nodes of the adjacent hidden layers. We also employed a rectified linear unit (ReLU, \citealt{relu}) activation function with `He uniform' \citep{he_weights} weight initialisation for each hidden layer in the encoder network, bottleneck and decoder network, whilst a linear activation function with `Glorot uniform' \citep{glorot_weights} weight initialisation was used for the output layer of the decoder network. In addition, we initialised all biases to zeros. It should be noted that we utilised the \textsc{Keras} deep learning framework \citep{keras} to construct the AE.}
    \label{fig:ae_architecture}
\end{figure}

We employed an unsupervised deep learning algorithm, known as an autoencoder (AE, \citealt{autoencoders}), as the first stage of our background subtraction model. Our overall objective for using an AE is to train it to learn to accurately reconstruct input data. The mechanism behind the AE can be separated into three main stages, that are known as the encoder network, bottleneck and decoder network. The overall architecture for a typical AE is shown in Figure \ref{fig:ae_architecture}.

The encoder network is composed of fully-connected layers that are responsible for processing an input dataset by performing nonlinear transformations of the input data into a compressed representation. This is achieved by decreasing the number of nodes in the fully-connected layers as the size of the encoder network increases. The compression is maximised within the bottleneck, where the number of nodes in the bottleneck determines the amount of compression. The underlying objective of the bottleneck is to obtain the lowest dimensional representation of the data that captures the most generalisable aspects about the data. From which, the compressed data is then passed to the decoder network for reconstruction. This involves decompressing the compressed data back into its original input dimensionality by increasing the number of nodes in the fully-connected layers as the size of the decoder network increases. If the AE is properly trained, the reconstructed feature values should closely resemble the feature values of the input data. Overall, an AE can be considered as a type of dimensionality reduction-based algorithm since it focuses on reducing the dimensionality of the input data. However, we reconfigure its functionality from a dimensionality reduction-based algorithm into an outlier detection algorithm by also examining the differences between the reconstructed outputs and the input data. We note that the decoder network has the same but reversed architecture to the encoder network, where the number of nodes in the fully-connected layers increases rather than decreases as the size increases.

In order to train the AE to generate accurate reconstructions, we used the mean squared error as our loss function. This measured the similarity between all of the input and reconstructed feature values of galaxies via the following equation:

\begin{equation}
	\text{Mean Squared Error} = \frac{1}{n}\sum_{i=1}^{n}(y_{i} - \hat{y}_{i})^{2} \ ,\label{eq:1}
\end{equation}

where $n$ is the number of input features, $y$ is the input feature values and $\hat{y}$ is the reconstructed feature values. 

We set the batch size, learning rate, optimiser algorithm\footnote[9]{We recommend the reader to refer to \cite{optimiser_algorithms} for an overview of different optimiser algorithms.} and architecture layout\footnote[10]{We considered the number of nodes in the bottleneck to be the most significant component of an AE's architecture since it is influential in the amount of generalization learned. We decided that the number of nodes in the bottleneck should be a tunable hyper-parameter whilst the number of nodes for the hidden layers in the encoder and decoder networks would remain fixed.} to be tunable hyper-parameters, where the full hyper-parameter search space is shown in Table \ref{tab:ae_hyperparameters}.

\begin{table*}
\centering
	\begin{tabular}{|p{4cm}|p{12cm}|}
		\hline
	    \hfil \textbf{Tunable hyper-parameter name} & \hfil \textbf{Hyper-parameter search space} \\
		\hline
		{\par\centering Batch size \par} & {\par\centering 256 or 512 or 1024 or 2048 \par} \\
		\hline
		{\par\centering Learning rate\par} & {\par\centering 0.0001 or 0.001 or 0.01 or 0.1 \par} \\
		\hline
	    {\par\centering Optimiser algorithm \par} & {\par\centering Adaptive Moment Estimation (Adam) or Adaptive Delta (Adadelta) or Adaptive Gradient Optimiser (Adagrad) or Adam Based On The Infinity Norm (Adamax) or Adam With Nesterov Momentum (Nadam) or Stochastic Gradient Descent (SGD) or Root Mean Squared Propagation (RMSprop) \par} \\
	    \hline
	    {\par\centering Architecture layout (number of nodes and hidden layers in the encoder network and bottleneck) \par} & {\par\centering \textbf{1} (13 nodes in first hidden layer, 11 nodes in second hidden layer, 9 nodes in third hidden layer, 7 nodes in the fourth hidden layer and 1 node in the bottleneck) or \textbf{2} (13 nodes in first hidden layer, 11 nodes in second hidden layer, 9 nodes in third hidden layer, 7 nodes in the fourth hidden layer and 3 nodes in the bottleneck) or \textbf{3} (13 nodes in first hidden layer, 11 nodes in second hidden layer, 9 nodes in third hidden layer, 7 nodes in the fourth hidden layer and 5 nodes in the bottleneck) \par} \\
		\hline
	\end{tabular}
	\caption{This table contains a list of tunable hyper-parameters for the AE as well as the range of values that were explorable in the hyper-parameter space via random search. We also set a maximum of ten thousand trainable epochs as well as enabling early stopping of the model training if the validation loss had not decreased by $0.001$ over fifty epochs from the best observed validation loss. Furthermore, we again remind the reader that the encoder and decoder networks had reversed symmetrical designs, so we did not specify the number of nodes or hidden layers for the decoder network within this table.}
	\label{tab:ae_hyperparameters}
\end{table*}          

We employed a separate machine learning algorithm, known as logistic regression (see \cite{logistic_regression} for further details), as the second stage of our background subtraction model. This served to convert the outputs of the AE into class predictions. In particular, we used the known class labels as the target variable and the mean squared error between the input and reconstructed feature values as the input variable, where if an input was poorly reconstructed by the AE then the corresponding mean squared error will be large too. From which, the logistic regression algorithm determines whether a galaxy should be classified as a cluster or field galaxy (i.e. the galaxy class with the higher predicted probability) when given the mean squared error of each galaxy. In this work, we decided to use the defaulted hyper-parameter values for the logistic regression algorithm (N.B. without regularisation) in the \textsc{Scikit-Learn} \citep{scikit-learn} machine learning library since we primarily wanted to examine the influence of the AE in our background subtraction model. We expect that tuning the hyper-parameters for the logistic regression algorithm may slightly improve the overall predictive performance of our background subtraction model but this can be explored further in future work. It should be noted that the logistic regression algorithm minimised the following loss function during its training:

\begin{equation}
	\text{Log Loss} = -\frac{1}{n}\sum_{i=1}^{n}y_{i}log(p_{i}) + (1 - y_{i})log(1 - p_{i}) \ ,\label{eq:2}
\end{equation}

where $n$ is the number of inputs, $y$ is the true class value (i.e. either $0$ or $1$) and $p$ is the predicted probability (i.e. between $0$ and $1$) of being a galaxy class. This loss function measured the difference between predicted probability and true class value of galaxies.

We utilised a random search (see \cite{random_search} for further details) strategy to examine the predictive performance of our background subtraction model with different hyper-parameter combinations, where random search is a computationally efficient approach that does not need to examine every hyper-parameter combination. Instead, it considers that hyper-parameter optimization can be characterised by a Gaussian process, such that only a minority of hyper-parameter combinations are actually important. For example, we can assume that a randomly selected hyper-parameter combination has a ninety-five per cent probability of being situated within the top five per cent of all possible hyper-parameter combinations from the optimum after conducting only sixty iterations of random search. At the same time, we employed a Monte Carlo cross-validation (see \cite{logistic_regression} for further details) strategy to examine the variability of the predictive performance of our background subtraction model with different weight initialisations and dataset compositions. This involved repeated random sampling of new training, validation and test sets over ten iterations to measure the average predictive performance across the ten iterations. Ideally, we aimed to select a hyper-parameter combination that offered consistency and good predictive performance.

To determine the optimal hyper-parameter combination of our background subtraction model, we utilised the trapezium rule to compute the area under a precision-recall curve (AUCPR, \citealt{pr_curve}) for each hyper-parameter combination, which is based on the following set of equations:

\begin{equation}
	\text{Precision} = \frac{\text{TP}}{\text{TP} + \text{FP}} \ ,\label{eq:3}
\end{equation}

\begin{equation}
	\text{Recall} = \frac{\text{TP}}{\text{TP} + \text{FN}} \ ,\label{eq:4}
\end{equation}

\begin{equation}
	\text{AUCPR} = \int \text{Precision} \ d\left(\text{Recall}\right) \ ,\label{eq:5}
\end{equation}

where $\text{TP}$ is the number of correctly classified `actual' cluster galaxies, $\text{FP}$ is the number of incorrectly classified `actual' cluster galaxies and $\text{FN}$ is the number of incorrectly classified `actual' field galaxies. Briefly, this metric measured the proportion of predictions that were predicted as cluster galaxies as well as the proportion of `actual' cluster galaxies that were recovered across all class probability thresholds. It is ideal for assessing the predictive performance of a model that focuses on correctly identifying `rare' instances (i.e. when there is a class imbalance). The optimal hyper-parameter combination would maximise the AUCPR for galaxies in our validation set.  

Next, we determined the corresponding optimal class probability threshold when using the optimal hyper-parameter combination. This involved comparing the F1 score \citep{f1_score} yielded for each class probability threshold (i.e. from $0$ to $1$ with class probability threshold step sizes of $0.01$) via the following equation:

\begin{equation}
	\text{F1 Score} = 2\left(\frac{\text{Precision} \ \text{x} \ \text{Recall}}{\text{Precision} \ + \ \text{Recall}}\right) \ .\label{eq:6}
\end{equation}

This metric was similar to AUCPR in functionality except it only considered the predictive performance at a specific class probability threshold. The optimal class probability threshold would maximise the F1 score for galaxies in our validation set.  

Lastly, we determined the overall classification accuracy of our background subtraction model at distinguishing between cluster and field galaxies in our test set. This involved computing the balanced accuracy \citep{balanced_accuracy} when using the optimal class probability threshold and optimal hyper-parameter combination via the following equation:

\begin{equation}
	\text{Balanced Accuracy} = \frac{1}{2}\left(\frac{\text{TP}}{\text{TP} + \text{FN}} + \frac{\text{TN}}{\text{TN} + \text{FP}}\right) \ ,\label{eq:7}
\end{equation}

where $\text{TP}$ is the number of correctly classified `actual' cluster galaxies, $\text{TN}$ is the number of correctly classified `actual' field galaxies, $\text{FP}$ is the number of incorrectly classified `actual' cluster galaxies and $\text{FN}$ is the number of incorrectly classified `actual' field galaxies. The primary advantage of using balanced accuracy rather than conventional classification accuracy is that balanced accuracy takes into account class imbalance whereas conventional classification accuracy assumes equal class sizes when measuring the predictive performance of a binary classification model.

\subsection{Establishing a scaling relation to estimate $r_{200}$}
\label{sec:scaling_relation_r200} 

It is beneficial to measure the richness of clusters within a characteristic radius (e.g. $r_{200}$, $r_{500}$, $r_{2500}$) because it enables a more straightforward comparison of cluster richness across different cluster catalogues. We decided to establish a scaling relation that predicts values for the characteristic radius of cross-matched WH15 and redMaPPer clusters from \S\S\ref{sec:prepare_photometry_for_background_subtraction}. We note that WH15 used a friend-of-friend grouping algorithm on galaxies with known spectroscopic or photometric redshifts to identify clusters in the Sloan Digital Sky Survey Data Release 12 (SDSS-III DR12, \citealt{sdss_dr12}) whilst redMaPPer used a red-sequence fitting algorithm on galaxies within colour-magnitude space to identify clusters in the Sloan Digital Sky Survey Data Release 8 (SDSS-III DR8, \citealt{sdss_dr8}). Subsequently, we obtained a total of $6064$ cross-matched WH15 and redMaPPer clusters between a redshift range of $0.1 \leq z \leq 0.35$. We decided to use these clusters because they were found via two conventional approaches for cluster detection. This enabled us to directly compare the consistency of richness estimates from using our novel cluster galaxy identification technique versus other cluster galaxy identification techniques. In this proof-of-concept study, we choose to employ only a subsample of $1000$ cross-matched WH15 and redMaPPer clusters when creating our scaling relation for time efficiency.

We also decided to use $r_{200}$ as our characteristic radius since the cluster galaxies in our training and test sets from the AMF11 catalogue were originally sampled within $r_{200}$. In particular, we utilised $r_{200}$ values that were estimated by WH15 as the dependent variable in our scaling relation, where their $r_{200}$ estimates were computed via a scaling relation between $r_{200}$ measurements from X-ray/weak lensing observations and total luminosity in the $r$ band of all the identified cluster galaxies. For this work, we assumed that these $r_{200}$ values can be considered as `actual' $r_{200}$ values. In Figure S5 (available online), we noticed that there was a strong linear relationship between WH15\footnote[11]{We refer to the $R_{L*}$ variable from the WH15 catalogue as WH15 richness, where they computed cluster richness by measuring the total luminosity of identified galaxy members as a function of the typical luminosity of galaxies in the $r$ filter.} and redMapper richness\footnote[12]{We refer to the $\lambda/S$ variable from the redMaPPer catalogue as redMaPPer richness, where they computed cluster richness by determining an expected richness which would yield the observed projected density, $i$ filter magnitudes and multiple colour indices of the identified red-sequence galaxies.}. This means that we can directly compare our predicted richnesses with the richness estimates of redMaPPer. Furthermore, we noticed that there was a non-linear relationship between $r_{200}$ and both WH15 and redMaPPer richnesses which was in accordance with the empirical richness-size relation observed in \cite{richness_size_relation}, where our `actual' $r_{200}$ values appeared to have greater variability at lower richnesses. As such, we expected that our scaling relation would have greater variability in $r_{200}$ at lower richnesses too.

\begin{figure}
\centering
	\includegraphics[width=\linewidth]{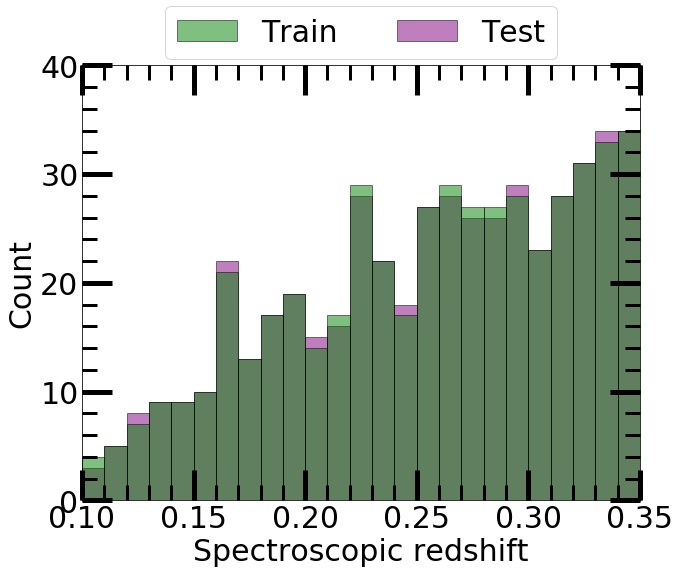}
	\includegraphics[width=\linewidth]{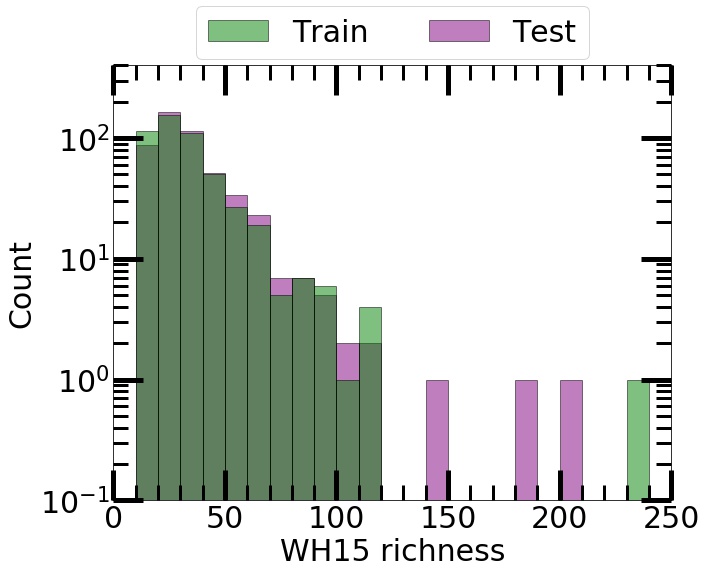}
    \caption{This figure shows histograms of the cluster spectroscopic redshift (top image) and WH15 richness (bottom image) distributions of clusters in our CMWR training (green) and test (purple) sets that were between a redshift range of $0.1 \leq z \leq 0.35$.}
    \label{fig:cmwr_spectroscopic_cluster_members_histogram}
\end{figure}

We then partitioned the cross-matched WH15 and redMaPPer clusters into a training set and test set. We nominally referred to these sets as the CMWR (i.e. cross-matched WH15 and redMaPPer) training and test sets to avoid confusion with the training and test sets created in \S\S\ref{sec:prepare_photometry_for_background_subtraction}. The purpose of having the CMWR training set was to determine the best fit coefficients of our scaling relation whilst the purpose of having the CMWR test set was to measure the predictive performance of our learned scaling relation. Since we knew the spectroscopic redshift of the CMWR clusters, we segmented them into fixed redshift bin sizes of $0.01$. This ensured that our training and test sets contained clusters from across the redshift scale via stratified sampling\footnote[13]{Stratified sampling is a strategy that minimises selection bias by splitting a dataset into new distributions that approximately resemble the original distribution.}. This involved randomly allocating approximately half of the clusters within each redshift bin into both sets, which resulted in our CMWR training and test sets containing $500$ clusters each. The spectroscopic redshift and richness distributions of clusters in our CMWR training and test sets can be seen in Figure \ref{fig:cmwr_spectroscopic_cluster_members_histogram}.

Next, we applied a search radius of $2.5$ Mpc at each cluster's spectroscopic redshift as well as reapplying the same observing flags mentioned in \S\S\ref{sec:prepare_photometry_for_background_subtraction} to acquire galaxies from SDSS-IV DR16. This gave us a total of $2020690$ galaxies, where our CMWR training set consisted of $1005167$ galaxies and our CMWR test set consisted of $1015523$ galaxies. We then applied our background subtraction model and colour-magnitude boundaries to count the number of cluster galaxies within each cluster. We established a linear scaling relation that was based on the number of identified cluster galaxies as an independent variable and $r_{200}$ as the dependent variable. This involved learning the best fit coefficients by minimising the residual sum of squares between the dependent and independent variables in a linear regression algorithm from the \textsc{Scikit-Learn} machine learning library, where we again used the defaulted hyper-parameter values for the linear regression algorithm. We note that our cross-matched WH15 and redMaPPer cluster sample contained many clusters with low richness but only a few clusters with high richness. As such, we decided to assign the WH15 richness of each cluster as individual weights in the linear regression algorithm to minimise the effect of overfitting to potential outliers from clusters with low richness.

\subsection{Preparation of a photometric dataset to estimate individual cluster richnesses}
\label{sec:prepare_photometry_to_predict_richness} 

\begin{figure*}
\centering
	\includegraphics[width=0.49\textwidth]{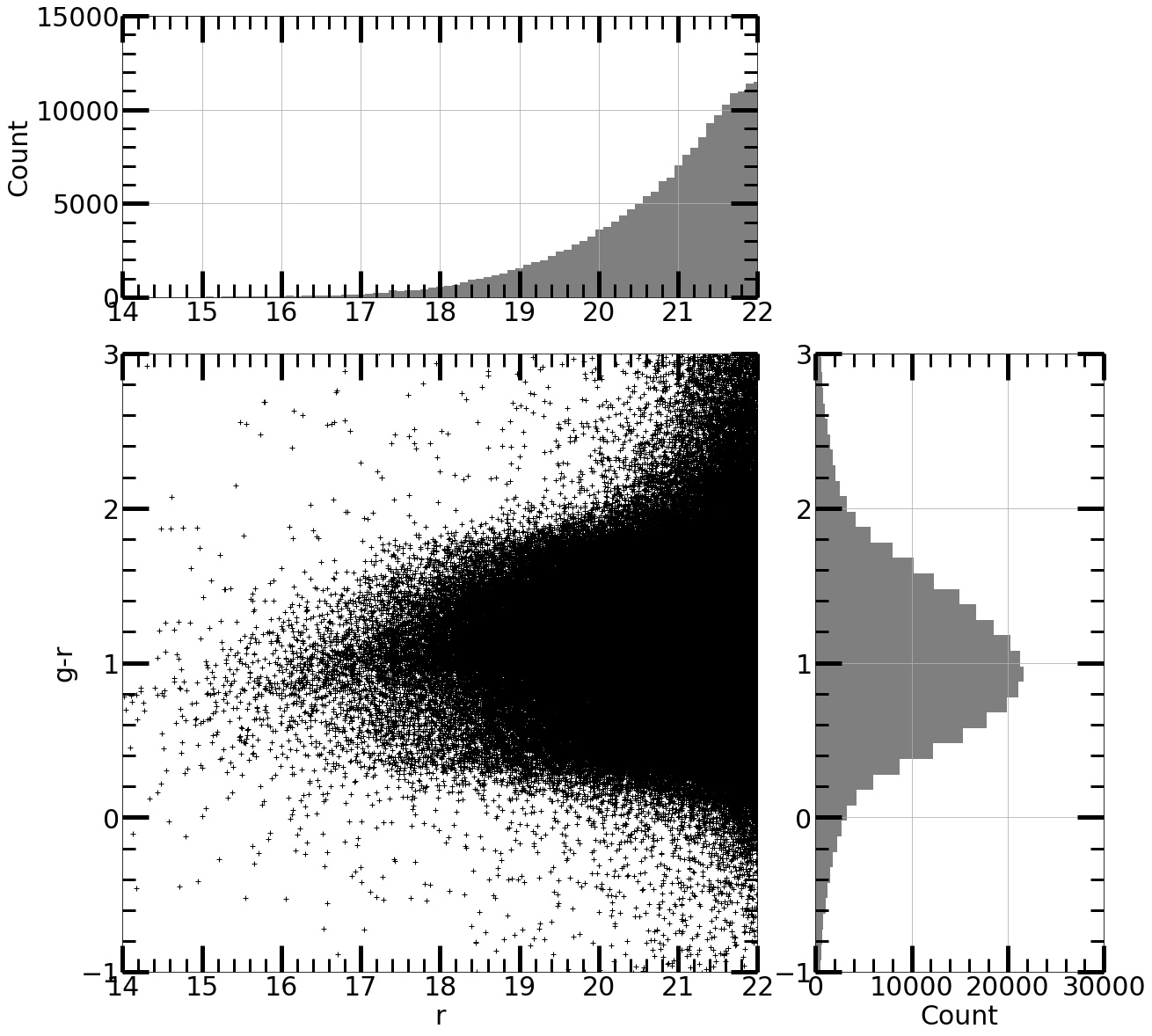}
	\includegraphics[width=0.49\textwidth]{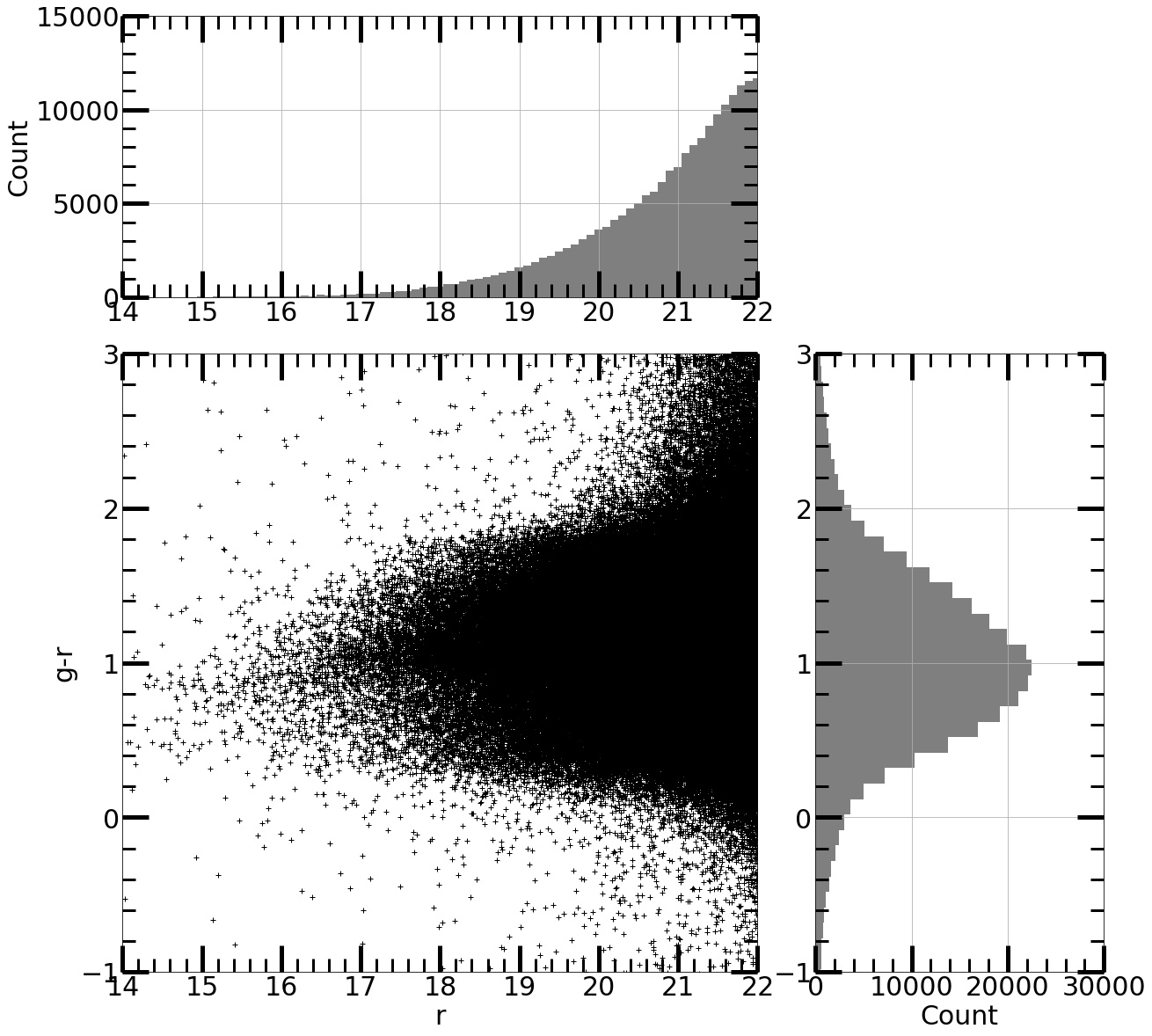}
    \caption{This figure shows colour-magnitude diagrams (using apparent magnitudes) of galaxies in our CMWR-$r_{200}$ training (left image) and test (right image) sets that were within an $r_{200}$ search radius and observed within SDSS-IV DR16.}
    \label{fig:cmwr_r200_cmd_cluster_vs_field}
\end{figure*}

In order to measure richness within $r_{200}$ of individual clusters, we first approximated $r_{200}$ for clusters in our CMWR training and test sets using the learned scaling relation from \S\S\ref{sec:scaling_relation_r200} to reacquire galaxies within $r_{200}$ from SDSS-IV DR16. We nominally referred to these new sets as the CMWR-$r_{200}$ training and test sets to avoid confusion with the CMWR training and test sets created in \S\S\ref{sec:scaling_relation_r200}. Similar to before, the purpose of having the CMWR-$r_{200}$ training set was to determine the best fit coefficients of a luminosity distribution fitting function whilst the purpose of having the CMWR-$r_{200}$ test set was to measure the predictive performance of the learned luminosity distribution fitting function. We obtained a total of $299807$ galaxies in our CMWR-$r_{200}$ training set and $306953$ galaxies in our CMWR-$r_{200}$ test set. The resultant color-magnitude diagrams of galaxies in our CMWR-$r_{200}$ training and test sets is shown in Figure \ref{fig:cmwr_r200_cmd_cluster_vs_field}.

\subsection{Using a luminosity distribution fitting function to estimate individual cluster richnesses within $r_{200}$}
\label{sec:luminosity_distribution_fitting}

We adopted a similar approach to the methodology described in \cite{schechter_function} to estimate the richness of individual clusters. \cite{schechter_function} showed that it was possible to use a luminosity distribution fitting function (i.e. the Schechter function) to do this. Briefly, this involved fitting the function to a composite luminosity distribution of cluster galaxies in order to determine best fit parameter values of the function. Then \cite{schechter_function} assumed that the best fit parameter values for $M^{*}$ and $\alpha$ can be applied universally to the luminosity distribution of individual clusters to locally fit for $n^{*}$ and thus estimate cluster richness. The Schechter function is expressed via the following equation:

\begin{equation}
	n(M)dM = [0.4ln(10)]n^{*}[10^{0.4(M^{*}-M)}]^{\alpha+1} e^{-10^{0.4(M^{*}-M)}}dM \ ,\label{eq:8}
\end{equation}

where $M$ is absolute magnitude, $n^{*}$ is the number of galaxies per unit magnitude, $M^{*}$ is the `characteristic' magnitude at which the distributions of faint and bright galaxies rapidly changes and $\alpha$ is the faint end slope parameter that describes the distribution of galaxies fainter than $M^{*}$. We note that $M^{*}$ and $\alpha$ directly influence the steepness of the bright and faint ends in the Schechter function whilst $n^{*}$ varies based on the observed number of galaxies within magnitude bins.

Firstly, we applied our background subtraction model and colour-magnitude boundaries to identify cluster galaxies from the CMWR-$r_{200}$ training set. We then performed Chi-squared fitting\footnote[14]{We used the \textit{curve fit} function from the \textsc{SciPy} Python library \citep{scipy} to perform Chi-squared fitting of the Schechter function, which returned the best fit parameter values that minimised the Chi-squared fitting error and also returned an estimated covariance matrix of the best fit parameter values.} with initialisation bounds for $M^{*}$ (i.e. between $-30$ and $-15$), $n^{*}$ (i.e. between $0$ and positive infinity) and $\alpha$ (i.e. between $-2$ and $-1$) when fitting the Schechter function to a composite luminosity distribution that consisted of a subsample of identified cluster galaxies which appeared to have high completeness (i.e. greater than $90$ per cent when using a base-10 logarithmic scale for the counts) between a restricted $r$ filter absolute magnitude (i.e. between $-25$ and $-21.5$) and redshift (i.e. between $0.1$ and $0.15$) range. At the same time, we also explored various $r$ filter absolute magnitude bin sizes (i.e. from $0.01$ to $3$ with step sizes of $0.01$) to obtain an optimal $r$ filter absolute magnitude bin size that minimised the Chi-squared fitting error and yielded galaxies across five or more $r$ filter absolute magnitude bins. Furthermore, we approximated the uncertainty in the number of identified cluster galaxies within each magnitude bin by assuming that the uncertainty followed a Poisson sampling hypothesis\footnote[15]{A Poisson sampling hypothesis assumed that the distribution of galaxies is dictated by a Poisson process, such that the standard deviation of the counts within each magnitude bin was based on the square root of the count \citep{schechter_function}.} when fitting the Schechter function. From which, we determined an optimal absolute magnitude bin size and best fit parameter values for $M^{*}$, $n^{*}$ and $\alpha$, where we also assumed that the best fit parameter values for $M^{*}$ and $\alpha$ can be applied universally to the luminosity distribution of individual clusters. 

We remind the reader that our background subtraction model had not yet been corrected for the incompleteness of faint galaxies from observing limitations. This meant that we had to derive completeness corrections for the luminosity distribution (i.e. using $r$ filter absolute magnitudes) of individual clusters at different redshifts. Initially, we grouped the identified cluster galaxies from the CMWR-$r_{200}$ training set into redshift intervals of $\pm 0.04(1 + z)$ that were centered in redshift bins from $0.105$ to $0.345$ with step sizes of $0.01$, where identified cluster galaxies from different redshifts can go into multiple bins. Next, we fitted a $100$ per cent completeness line across adjacent $r$ filter apparent magnitude bins\footnote[16]{We used an $r$ filter apparent magnitude bin size that corresponded to the optimal $r$ filter absolute magnitude bin size. In addition, when working with the luminosity distribution of individual clusters we only considered cluster galaxies within an $r$ filter absolute magnitude range of $-25$ to $-20.5$, where $-20.5$ was the $r$ filter absolute magnitude limit that was used to determine WH15 richness in \cite{wh15}.} that were on the bright side of the peak and within the completeness limit of the AMF11 catalogue for each redshift interval. We then approximated the completeness fraction of the faintest $r$ filter apparent magnitude bin (N.B. we considered the two faintest magnitude bins on the bright side of the peak beyond $z > 0.14$ and three faintest magnitude bins on the bright side of the peak beyond $z > 0.33$ as the incompleteness of galaxies became more visibly noticeable for more magnitude bins at higher redshifts) by calculating the fraction in the expected number of cluster galaxies (i.e. based on the $100$ per cent completeness line) to the observed number of cluster galaxies (i.e. identified by our background subtraction model).

We applied these completeness fractions to the luminosity distribution of individual clusters by multiplying the observed count of the faintest (N.B. we again considered the two faintest magnitude bins on the bright side of the peak beyond $z > 0.14$ and three faintest magnitude bins on the bright side of the peak beyond $z > 0.33$) $r$ filter absolute magnitude bin\footnote[17]{Since our completeness fractions were measured in $r$ filter apparent magnitudes, we had to convert between $r$ filter apparent magnitudes and $r$ filter absolute magnitudes to determine the relevant completeness fraction.} on the bright side of the peak by the completeness fraction of the corresponding $r$ filter apparent magnitude bin within the nearest redshift interval. We then replaced the uncertainty range of the observed count in the magnitude bin with this computed completeness correction value as the new lower and upper uncertainty limits when performing Chi-square fitting. This ensured that the Schechter function did not fit to incomplete $r$ filter absolute magnitude bins.

Finally, we estimated cluster richnesses within $r_{200}$ by integrating\footnote[18]{We utilised the incomplete Gamma function (see Equation $27$ in \cite{schechter_function}) to compute the integral.} the locally fit Schechter function. This gave us the expected number of cluster galaxies within $r_{200}$ that had an $r$ filter absolute magnitude brighter than $-20.5$. We also compared our estimated cluster richnesses with WH15 richnesses, spectroscopic redshift, `actual' $r_{200}$ and redMaPPer richnesses in order to examine the predictive performance of the optimal $r$ filter absolute magnitude bin size and best fit parameters for $M^{*}$ and $\alpha$ in the Schechter function. We note that WH15 richness was specific to $r_{200}$ whereas redMaPPer richness was specific to redMaPPer's own scaling radius rather than $r_{200}$. This meant that we could directly quantify the error between our estimated cluster richnesses and WH15 richnessses by using root mean squared error as a metric.

\section{Results}
\label{sec:results}

\subsection{Model tuning analyses}
\label{sec:model_tuning_analyses}

\subsubsection{Analysis of our trained background subtraction model}
\label{sec:ae_hyper-parameter_analysis}

We conducted ten iterations of Monte Carlo cross-validation to measure the variability of the predictive performance of our background subtraction model, as well as conducting sixty iterations of random search on the tunable hyper-parameters of our background subtraction model, to determine an optimal hyper-parameter combination that maximised the AUCPR of galaxies in our validation set. It can be seen in Table S1 (available online) that the optimum hyper-parameter combination was as follows: optimal batch size = $2048$; optimal learning rate = $0.0001$; optimal optimiser algorithm = RMSprop and optimal architecture layout = $3$. This optimum hyper-parameter combination yielded a mean AUCPR value of $40.24$ per cent with a standard deviation of $1.85$ per cent for galaxies in our validation set. Furthermore, it can be seen in Table S2 (available online) that the optimum class probability threshold was $0.29$, when using the optimum hyper-parameter combination. This optimum class probability threshold yielded a F1 score of $48.92$ per cent for galaxies in our validation set.

\subsubsection{Analysis of our established scaling relation to estimate $r_{200}$}
\label{sec:scaling_relation_analysis}

\begin{figure*}
\centering
	\includegraphics[width=\linewidth]{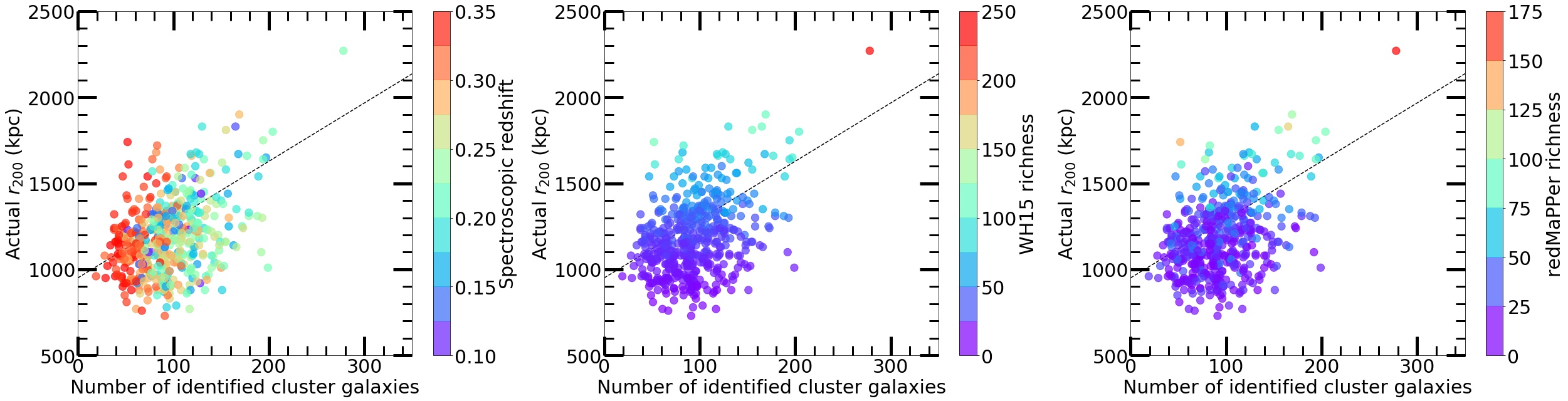}
	\caption{This figure shows our scaling relation (black dotted line) to estimate the $r_{200}$ of clusters. It used the `actual' $r_{200}$ of clusters from our CMWR training set as a dependent variable and the number of cluster galaxies identified by our background subtraction model within a $2.5$ Mpc search radius at each cluster's spectroscopic redshift as an independent variable. We also display the corresponding spectroscopic redshift (left image), WH15 richness (middle image) and redMaPPer richness (right image) of each cluster.}
    \label{fig:train_scaling_r200}
\end{figure*}

\begin{figure}
\centering
	\includegraphics[width=\linewidth]{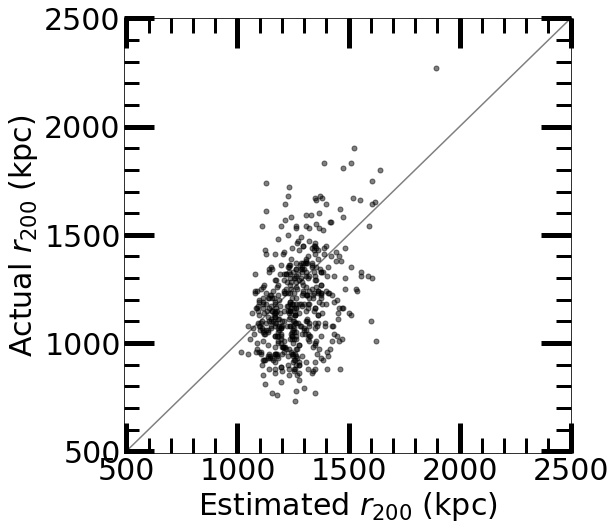}
	\caption{This figure shows a direct comparison of $r_{200}$ predicted by our scaling relation with the `actual' $r_{200}$ of clusters from our CMWR training set.}
    \label{fig:train_scaling_r200_pred_vs_actual}
\end{figure}

We constructed a scaling relation using clusters in our CMWR training set to estimate the $r_{200}$ of each cluster when given the number of cluster galaxies identified by our background subtraction model as an input. The best fit coefficients of our scaling relation were determined by minimising the weighted residual sum of squares between the independent and dependent variables, where our scaling relation is defined via the following equation: 

\begin{equation}
pred_{r_{200}} = (3.39 \pm 0.23) n_{gal} \ + \ (950.65 \pm 25.25) \ ,\label{eq:9}
\end{equation}

where $pred_{r_{200}}$ is the predicted $r_{200}$, $n_{gal}$ is the number of cluster galaxies identified within a $2.5$ Mpc search radius at each cluster's spectroscopic redshift and the uncertainty represents the standard error of the parameter estimates. In Figure \ref{fig:train_scaling_r200}, it can be seen that there was a larger drop in the number of cluster galaxies identified by our background subtraction model at higher redshifts (i.e. $z > 0.3$) when compared to the number of identified cluster galaxies at lower redshifts with the same `actual' $r_{200}$ values. This was likely due to cluster galaxies at higher redshifts having larger observed photometric errors, which made it more difficult for our background subtraction model to identify these cluster galaxies. We note that we obtained a Pearson correlation coefficient value of $0.39$ between the number of identified cluster galaxies and `actual' $r_{200}$ variables. We also observed that both WH15 and redMaPPer richnesses appeared to somewhat linearly increase with `actual' $r_{200}$ and the number of identified cluster galaxies. Furthermore, in Figure \ref{fig:train_scaling_r200_pred_vs_actual}, we compared the predictive performance of our predicted $r_{200}$ with the `actual' $r_{200}$, where we found that our predicted $r_{200}$ was quite comparable to the `actual' $r_{200}$ across all cluster sizes. Although, we noticed there was greater variability in the predicted $r_{200}$ at lower cluster richnesses, where we obtained a root mean squared error of $218.14$ and a median absolute percentage error of $11.89$ per cent between our predicted and `actual' $r_{200}$ values.

\subsubsection{Analysis of the best fit parameters for a luminosity distribution fitting function to estimate individual cluster richnesses Within $r_{200}$}
\label{sec:schechter_parameter_analysis}

\begin{figure}
\centering
	\includegraphics[width=\linewidth]{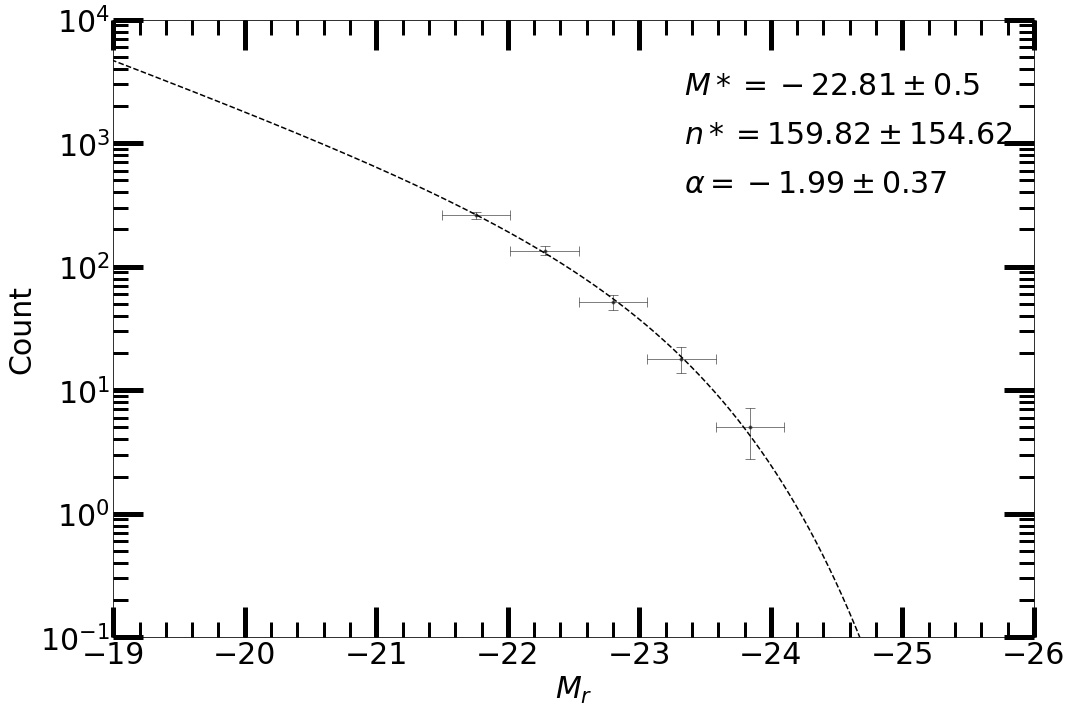}
	\caption{This figure shows the best fit Schechter function (black dotted line) overlaid on a composite luminosity distribution (using $r$ filter absolute magnitudes) that consisted of a subsample of identified cluster galaxies from our CMWR-$r_{200}$ training set with an optimal $r$ filter absolute magnitude bin size of $0.52$. The best fit parameter values and their respective standard deviations are displayed in the top right corner of the figure. The x-axis error bars display the width of each $r$ filter absolute magnitude bin and the y-axis error bars display the standard deviation of the observed count within each $r$ filter absolute magnitude bin when assuming a Poisson sampling hypothesis.}
    \label{fig:train_optimal_luminosity_distribution_pred_galaxies}
\end{figure}

We used a Chi-squared fitting approach to determine the best fit parameters of the Schechter function when fitting to a composite luminosity distribution that consisted of a subsample of identified cluster galaxies from our CMWR training set with high completeness. We also simultaneously determined an optimal $r$ filter absolute magnitude bin size that minimised the Chi-squared fitting error and yielded galaxies across five or more $r$ filter absolute magnitude bins. In Table S3 (available online), we identified an optimal $r$ filter absolute magnitude bin size of $0.52$ that had corresponding best fit parameter values of $M^{*}$ = $-22.81$ with a standard deviation of $\pm 0.5$; $n^{*}$ = $159.82$ with a standard deviation of $\pm 154.62$ and $\alpha$ = $-1.99$ with a standard deviation of $\pm 0.37$. In Figure \ref{fig:train_optimal_luminosity_distribution_pred_galaxies}, we display the composite luminosity distribution and fitted Schechter function using the optimal $r$ filter absolute magnitude bin size and best fit parameter values.

\begin{figure*}
\centering
	\includegraphics[width=\textwidth]{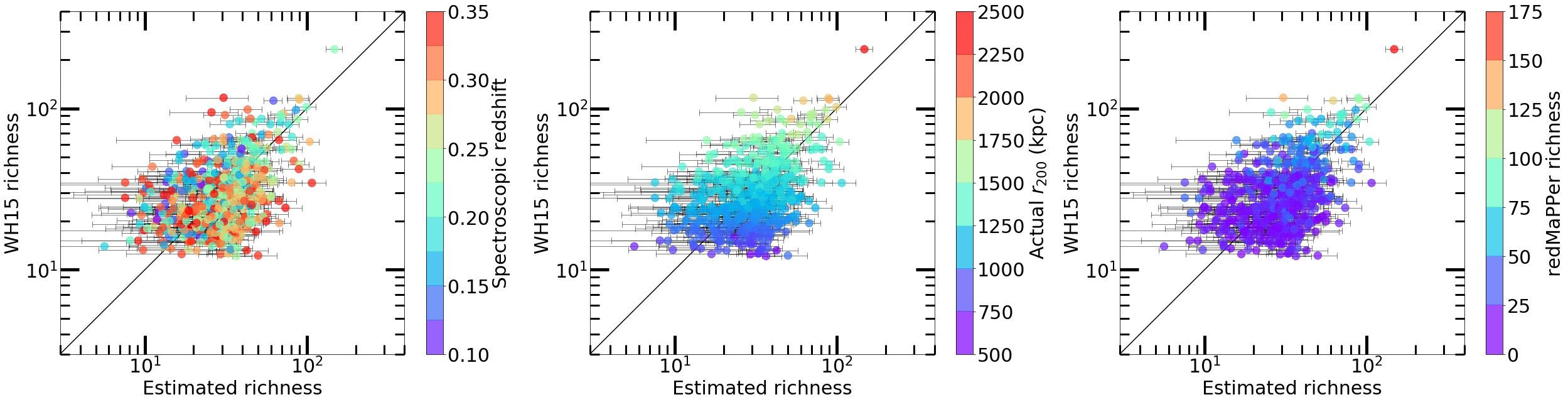}
	\caption{This figure shows a direct comparison between our estimated cluster richnesses and WH15 richnesses of clusters from our CMWR-$r_{200}$ training set when using the optimal $r$ filter absolute magnitude bin size and best fit parameter values for $M^{*}$ and $\alpha$. We also display the corresponding spectroscopic redshifts (left image), `actual' $r_{200}$ (middle image) and redMaPPer richness (right image) for each cluster. The x-axis error bars display the standard deviation of the locally fit $n^{*}$ when computing the integral of the Schechter function to determine our estimated cluster richnesses.}
    \label{fig:train_optimal_pred_richness}
\end{figure*}

We then used the optimal $r$ filter absolute magnitude bin size and best fit parameter values for $M^{*}$ and $\alpha$ to fit the Schechter function to the luminosity distribution of individual clusters from our CMWR-$r_{200}$ training set. This enabled us to estimate individual cluster richnesses by integrating the locally fit Schechter function. Subsequently, we obtained a root mean squared error of $18.06$ and a median absolute percentage error of $34.33$ per cent between our estimated cluster richnesses and WH15 richnesses within $r_{200}$. In Figure \ref{fig:train_optimal_pred_richness}, we noticed that WH15 richnesses had a strong linear correlation with our estimated cluster richnesses. We also observed that spectroscopic redshifts seemed to have no distinguishable correlation with our estimated cluster richnesses. In addition, we noticed that there was a strong linear correlation between `actual' $r_{200}$, redMaPPer richnesses and our estimated cluster richnesses. These results confirmed that our approach to estimate individual cluster richnesses was appropriate since we did not train any of our models to minimise cluster richness prediction error but we still obtained strong correlations with WH15 and redMaPPer richnesses. Furthermore, we were aware that our CMWR-$r_{200}$ training set contained clusters that were not truly unseen, as we had utilised these clusters before to create our scaling relation. Although, it was still interesting to test our methodology on clusters that were seen and unseen to compare differences in predictive performance.

\subsection{Overall performance analyses with test sets}
\label{sec:analyses_with_test_sets}

\begin{figure*}
\centering
	\includegraphics[width=0.49\linewidth]{Color_magnitude_diagrams_test_cluster_vs_field_galaxies.jpg}
	\includegraphics[width=0.49\linewidth]{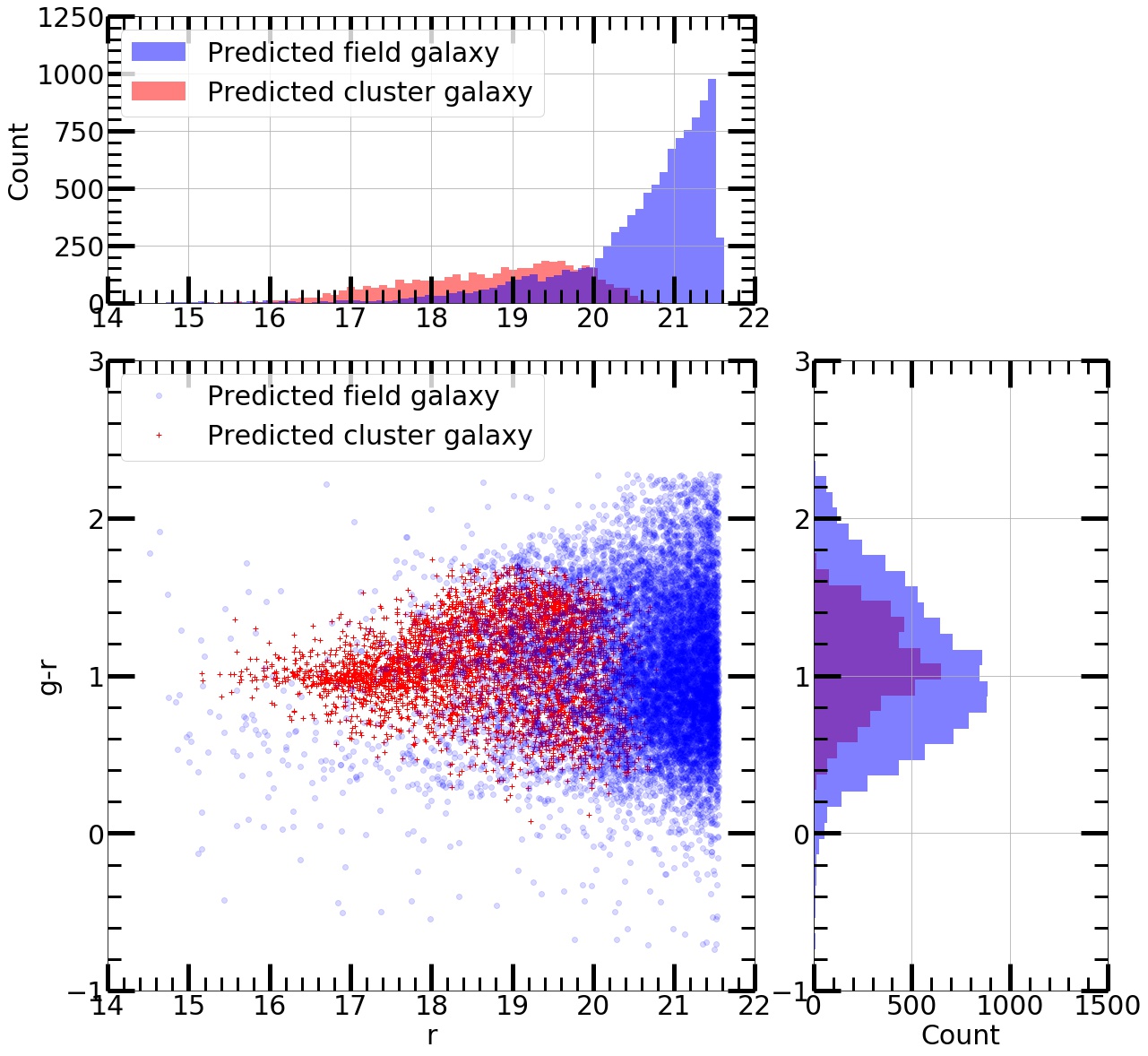}
	\caption{This figure shows a direct comparison of the colour-magnitude diagrams (using apparent magnitudes) for the `actual' (left image) and predicted (right image) cluster (red cross) and field (blue circle) galaxies in our test set.}
    \label{fig:cmd_actual_vs_predict_test}
\end{figure*}

\begin{figure*}
\centering
	\includegraphics[width=0.49\linewidth]{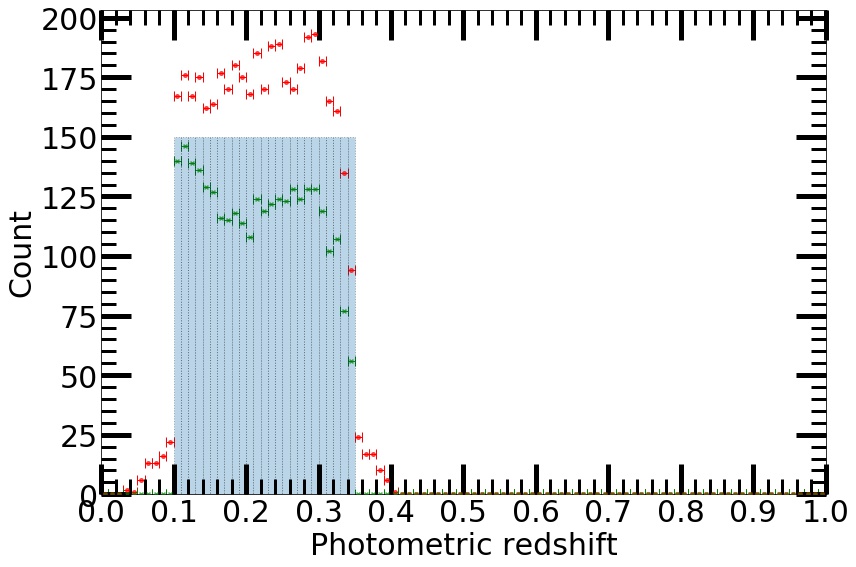}
	\includegraphics[width=0.49\linewidth]{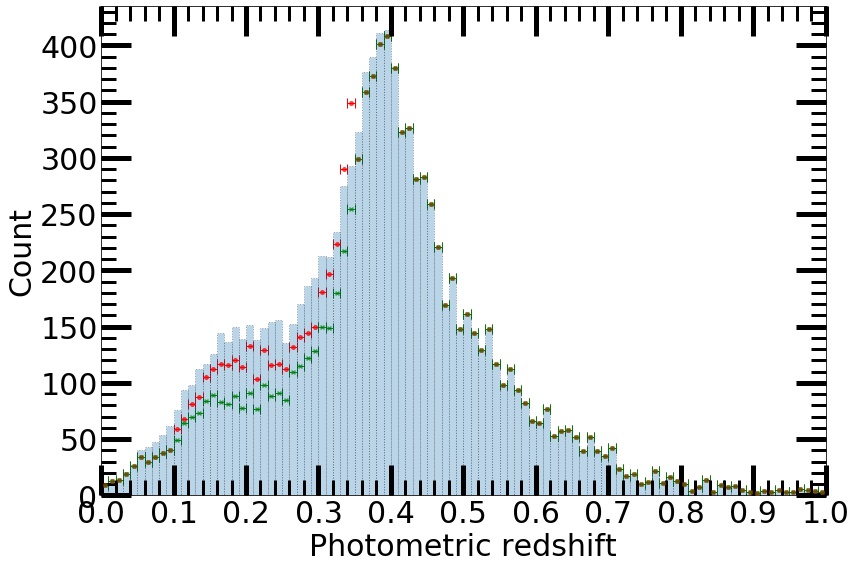}
	\caption{This figure shows histograms of the number of identified cluster (left image) and field (right image) galaxies in our test set when using fixed redshift bin sizes of $0.01$. The blue fill with black dotted lines represents the original number of `actual' cluster or field (N.B. we only display field galaxies that had an available photometric redshift) galaxies within each redshift bin. The red points represent the number of cluster or field galaxies identified by our background subtraction model within each redshift bin, the green crosses represent the number of `actual' cluster or field galaxies identified by our background subtraction model within each redshift bin and the x-axis error bars display the width of each redshift bin.}
    \label{fig:ae_recovered_redshift_test}
\end{figure*}

\begin{figure*}
\centering
	\includegraphics[width=0.49\linewidth]{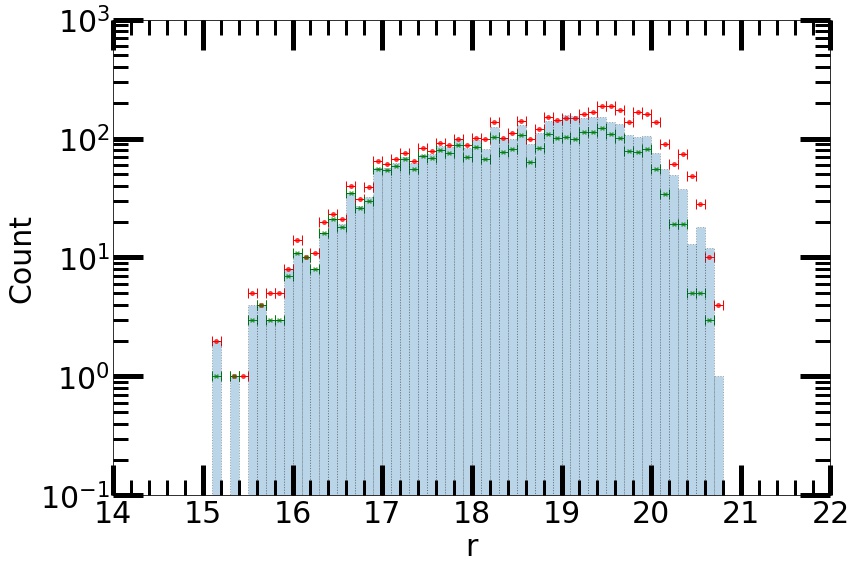}
	\includegraphics[width=0.49\linewidth]{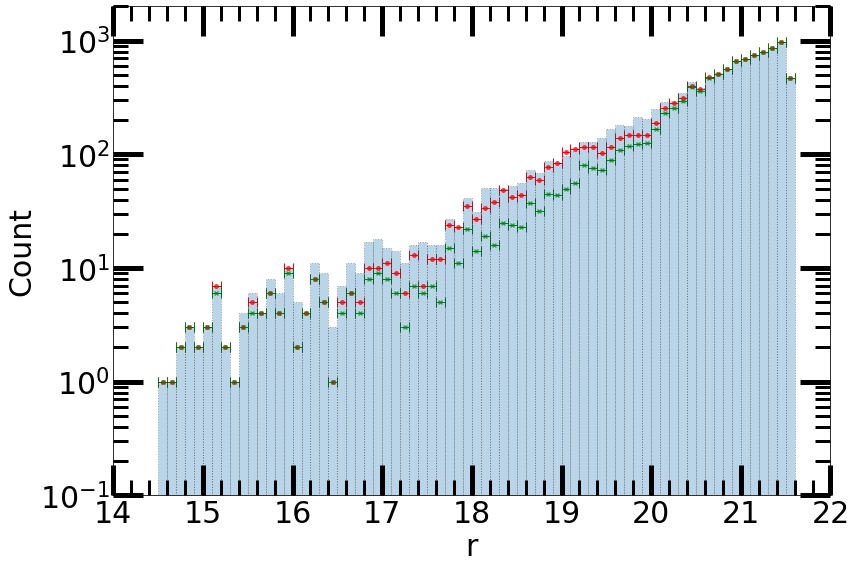}
	\caption{This figure shows histograms of the number of identified cluster (left image) and field (right image) galaxies in our test set when using fixed $r$ filter apparent magnitude bin sizes of $0.1$. The blue fill with black dotted lines represents the original number of `actual' cluster or field galaxies within each $r$ filter apparent magnitude bin. The red points represent the number of cluster or field galaxies identified by our background subtraction model within each $r$ filter apparent magnitude bin, the green crosses represent the number of `actual' cluster or field galaxies identified by our background subtraction model within each $r$ filter apparent magnitude bin and the x-axis error bars display the width of each $r$ filter apparent magnitude bin.}
    \label{fig:ae_recovered_rband_test}
\end{figure*}

\begin{figure*}
\centering
	\includegraphics[width=0.49\linewidth]{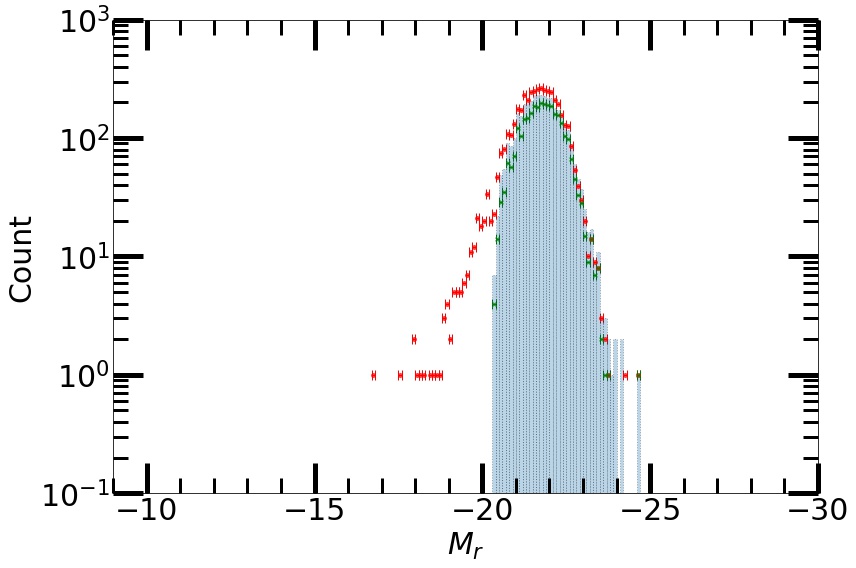}
	\includegraphics[width=0.49\linewidth]{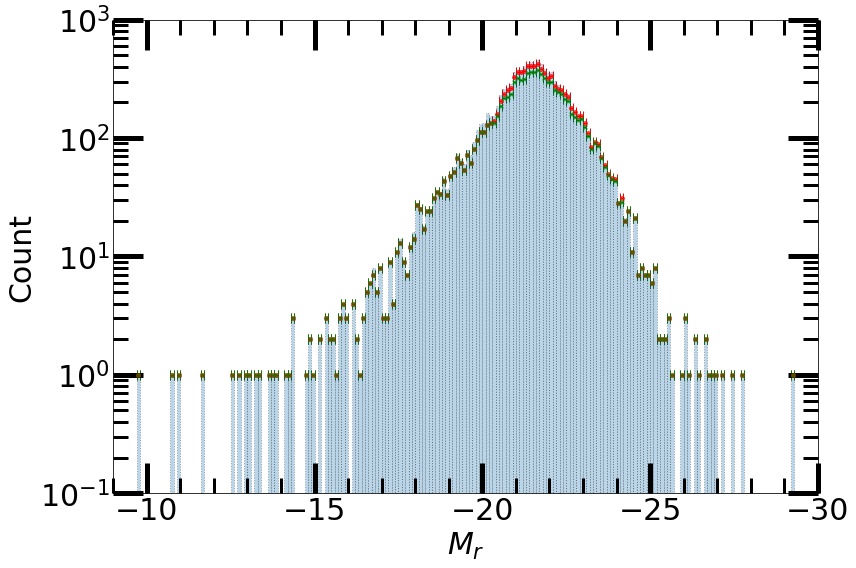}
	\caption{This figure shows histograms of the number of identified cluster (left image) and field (right image) galaxies in our test set when using fixed $r$ filter absolute magnitude bin sizes of $0.1$. The blue fill with black dotted lines represents the original number of `actual' cluster or field (N.B. we only display field galaxies that had an available photometric redshift) galaxies within each $r$ filter absolute magnitude bin. The red points represent the number of cluster or field galaxies identified by our background subtraction model within each $r$ filter absolute magnitude bin, the green crosses represent the number of `actual' cluster or field galaxies identified by our background subtraction model within each $r$ filter absolute magnitude bin and the x-axis error bars display the width of each $r$ filter absolute magnitude bin.}
    \label{fig:ae_recovered_abs_rband_test}
\end{figure*}

\begin{figure}
\centering
	\includegraphics[width=\linewidth]{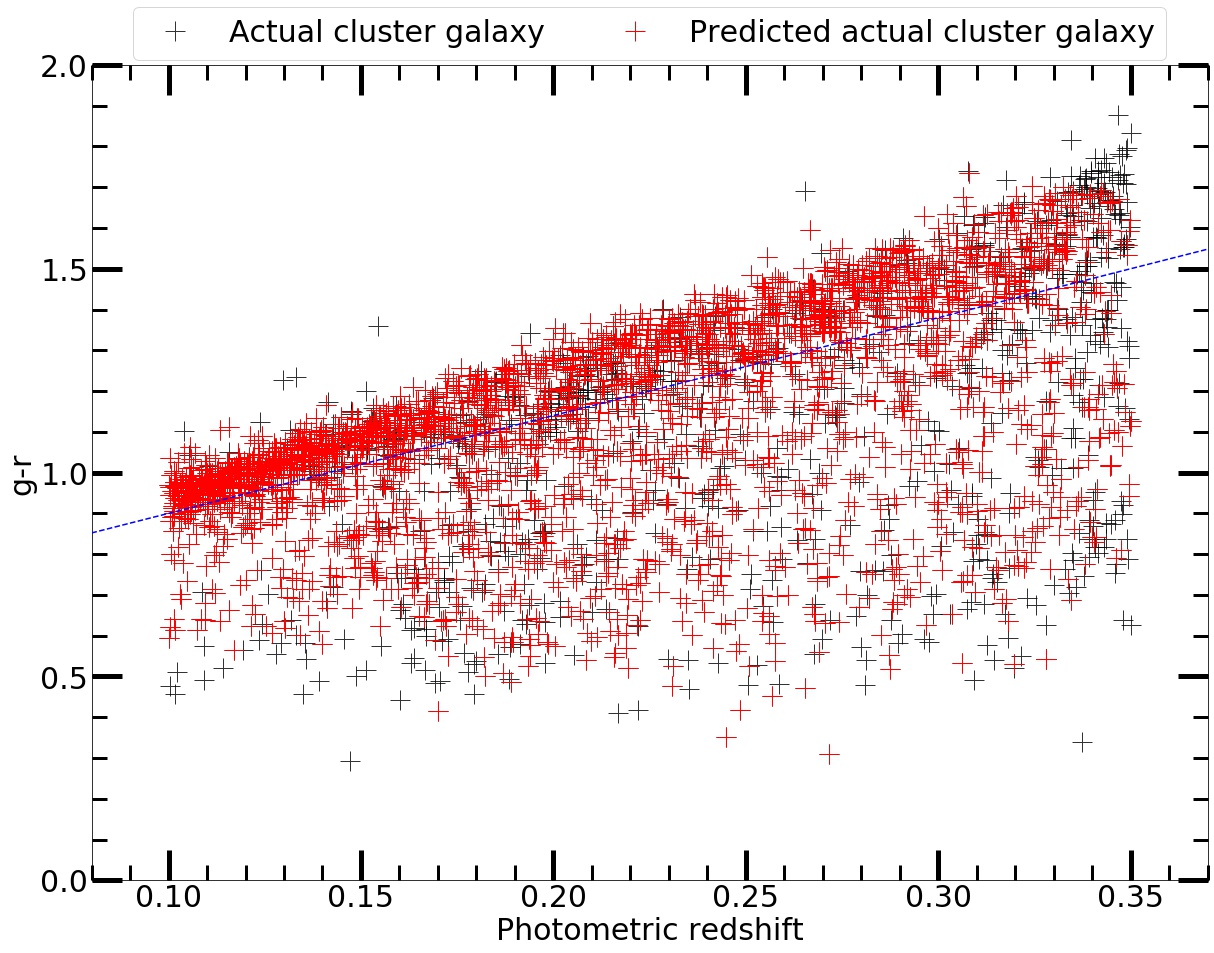}
	\caption{This figure shows a comparison of the `red' and `blue' `actual' cluster galaxies (black cross) in our test set that were identified (red cross) by our background subtraction model at different redshifts, where we assumed that galaxies above the blue dashed line were `red' and galaxies below the blue dashed line were `blue'.}
    \label{fig:ae_recovered_colour_test}
\end{figure}

We further assessed our entire methodology on clusters belonging to our various test sets to obtain an unbiased evaluation of the true predictive performance of our models. Firstly, we applied our background subtraction model to cluster and field galaxies in our test set. This yielded a F1 score of $72.81$ per cent and a balanced accuracy of $83.20$ per cent when using the optimal hyper-parameter combination and optimal class probability threshold for our background subtraction model. In Figure \ref{fig:cmd_actual_vs_predict_test}, we display a direct comparison of the `actual' and predicted cluster and field galaxies. It can be seen that our background subtraction model learned to correctly classify almost all of the field galaxies surrounding the cluster galaxies but it made more incorrect classifications in regions where the `actual' cluster and field galaxies had greater overlap within colour-magnitude space. Meanwhile, in Figure \ref{fig:ae_recovered_redshift_test}, we compared the number of cluster and field galaxies identified by our subtraction model across redshift bin sizes of $0.01$. At lower redshifts (i.e. $z \leq 0.45$), we noticed that our background subtraction model slightly underestimated (i.e. misclassified `actual' field galaxies as cluster galaxies or misclassified `actual' cluster galaxies as field galaxies) the overall number of `actual' cluster and field galaxies. In particular, we noticed a larger drop in the number of identified `actual' cluster galaxies between $0.3 \leq z \leq 0.35$, which was similar to our observation in Figure \ref{fig:train_scaling_r200}. Correspondingly, at higher redshifts (i.e. $z > 0.45$), we found that our background subtraction model correctly classified almost all of the galaxies. In Figures \ref{fig:ae_recovered_rband_test} and \ref{fig:ae_recovered_abs_rband_test}, we compared the number of cluster and field galaxies identified by our subtraction model across $r$ filter apparent and absolute magnitude bin sizes of $0.1$ respectively. In both magnitude distributions, we noticed that our background subtraction model slightly underestimated the overall number of `actual' cluster galaxies at all magnitudes. We also noticed that our background subtraction model slightly underestimated the overall number of `actual' field galaxies at intermediate brightnesses (i.e. between $16.5$ and $20.5$ in $r$ filter apparent magnitude and between $-24$ and $-20$ in $r$ filter absolute magnitude) but correctly classified almost all of the other fainter and brighter `actual' field galaxies. Furthermore, in Figure \ref{fig:ae_recovered_colour_test} we examined the proportion of `red' and `blue' `actual' cluster galaxies that were identified by our background subtraction model at different redshifts. We found that our background subtraction model identified $84.32$ per cent of `red' `actual' cluster galaxies and recovered $73.11$ per cent of `blue' `actual' cluster galaxies between a redshift range of $0.1 \leq z \leq 0.35$. This indicated that our background subtraction model was more confident at identifying `red' `actual' cluster galaxies than `blue' `actual' cluster galaxies, which was likely due to the `blue' `actual' cluster galaxies having greater overlap with field galaxies within colour-magnitude space.

\begin{figure*}
\centering
	\includegraphics[width=\linewidth]{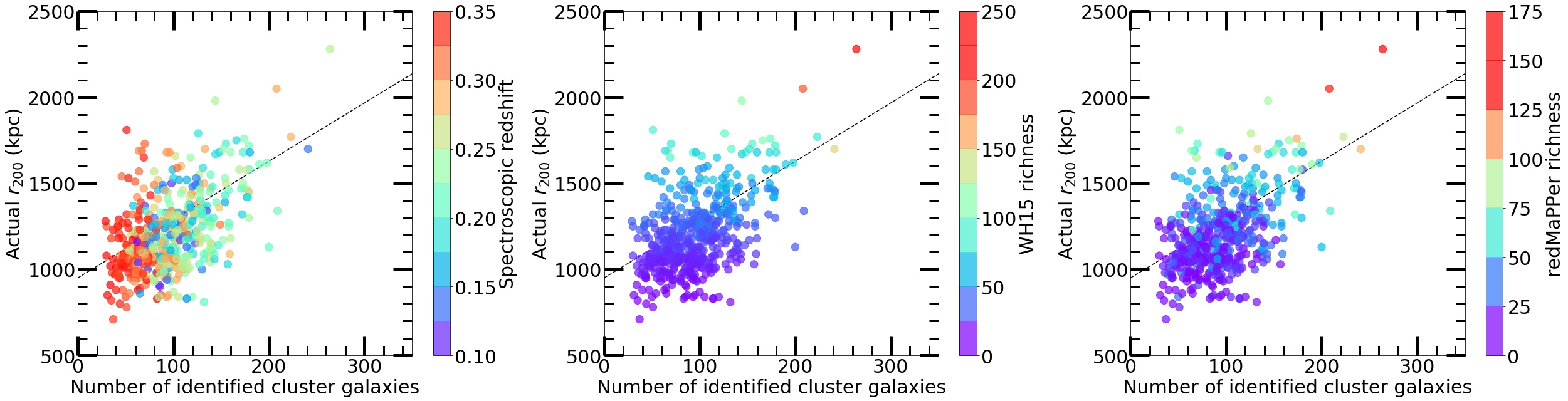}
	\caption{This figure is equivalent to Figure \ref{fig:train_scaling_r200} except we overlaid our learned scaling relation (black dotted line) on clusters in our CMWR test set.}
    \label{fig:test_scaling_r200}
\end{figure*}

\begin{figure}
\centering
	\includegraphics[width=\linewidth]{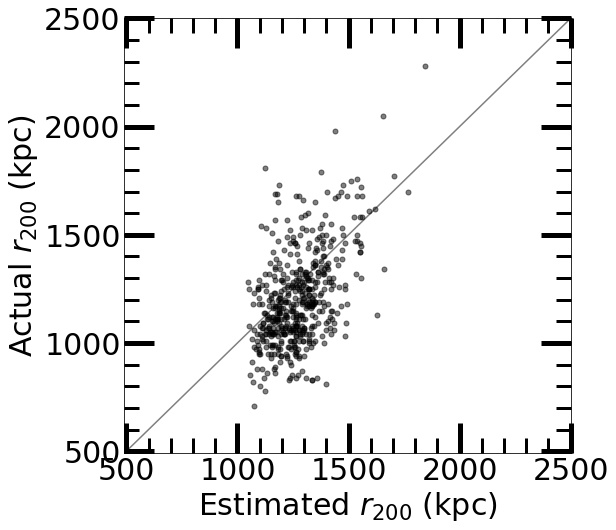}
	\caption{This figure is equivalent to Figure \ref{fig:train_scaling_r200_pred_vs_actual} except it compared the predicted and `actual' $r_{200}$ of clusters in our CMWR test set.}
    \label{fig:test_scaling_r200_pred_vs_actual}
\end{figure}

We then applied our learned scaling relation and colour-magnitude boundaries to clusters in our CMWR test set to approximate $r_{200}$ for each cluster. In Figure \ref{fig:test_scaling_r200}, we noticed that the number of identified cluster galaxies and `actual' $r_{200}$ was relatively consistent with our learned scaling relation from Figure \ref{fig:train_scaling_r200}. From which, we obtained a Pearson correlation coefficient value of $0.50$ between the number of identified cluster galaxies and `actual' $r_{200}$ variables in Figure \ref{fig:test_scaling_r200}. We also noticed that our predicted and `actual' $r_{200}$ values in Figure \ref{fig:test_scaling_r200_pred_vs_actual} was similar to the overall trend observed in Figure \ref{fig:train_scaling_r200_pred_vs_actual}, where we obtained a root mean squared error of $200.86$ and a median absolute percentage error of $11.66$ per cent between our predicted and `actual' $r_{200}$ values in Figure \ref{fig:test_scaling_r200_pred_vs_actual}.

\begin{figure*}
\centering
	\includegraphics[width=\textwidth]{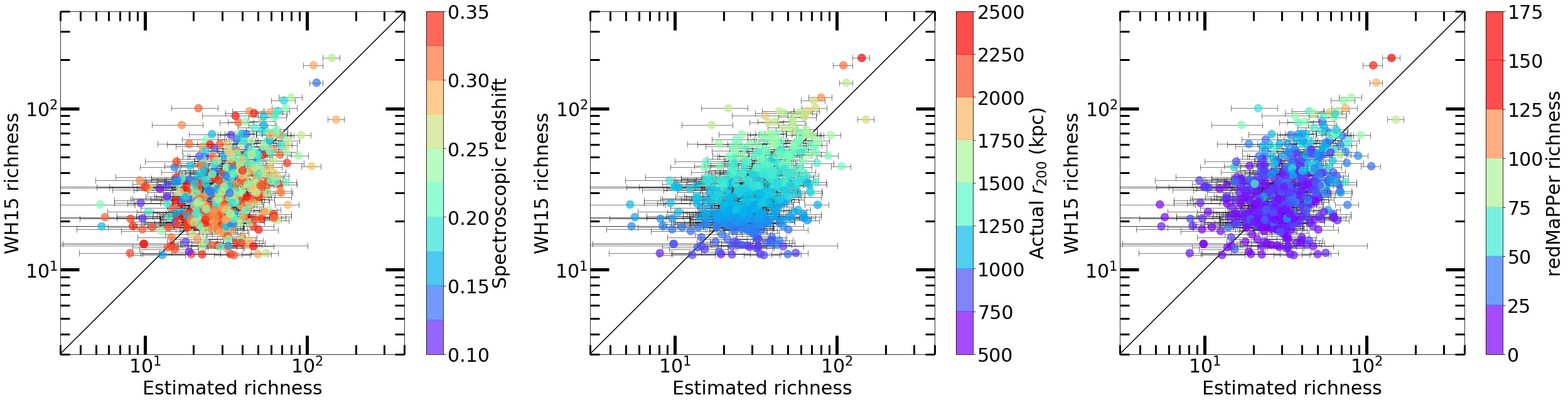}
	\caption{This figure is equivalent to Figure \ref{fig:train_optimal_pred_richness} except it was applied to unseen clusters from the CMWR-$r_{200}$ test set.}
    \label{fig:test_optimal_pred_richness}
\end{figure*}

Finally, we examined the predictive performance of the optimal $r$ filter absolute magnitude bin size and best fit parameter values for $M^{*}$ and $\alpha$ in the Schechter function on individual clusters in our CMWR-$r_{200}$ test set. In Figure \ref{fig:test_optimal_pred_richness}, we noticed that the overall trends between our estimated cluster richnesses and WH15 richnesses with spectroscopic redshifts, `actual' $r_{200}$ and redMaPPer richnesses were again consistent with Figure \ref{fig:train_optimal_pred_richness}, where our estimated cluster richnesses had no distinct correlation with spectroscopic redshifts and our estimated cluster richnesses linearly increased with `actual' $r_{200}$ and redMaPPer richnesses. Subsequently, we obtained a root mean squared error of $18.04$ and a median absolute percentage error of $33.50$ per cent between our estimated cluster richnesses and WH15 richnesses within $r_{200}$.

\subsection{Examining the importance of input features to our background subtraction model}
\label{sec:examining_feature_importance}

\begin{figure}
\centering
	\includegraphics[width=\linewidth]{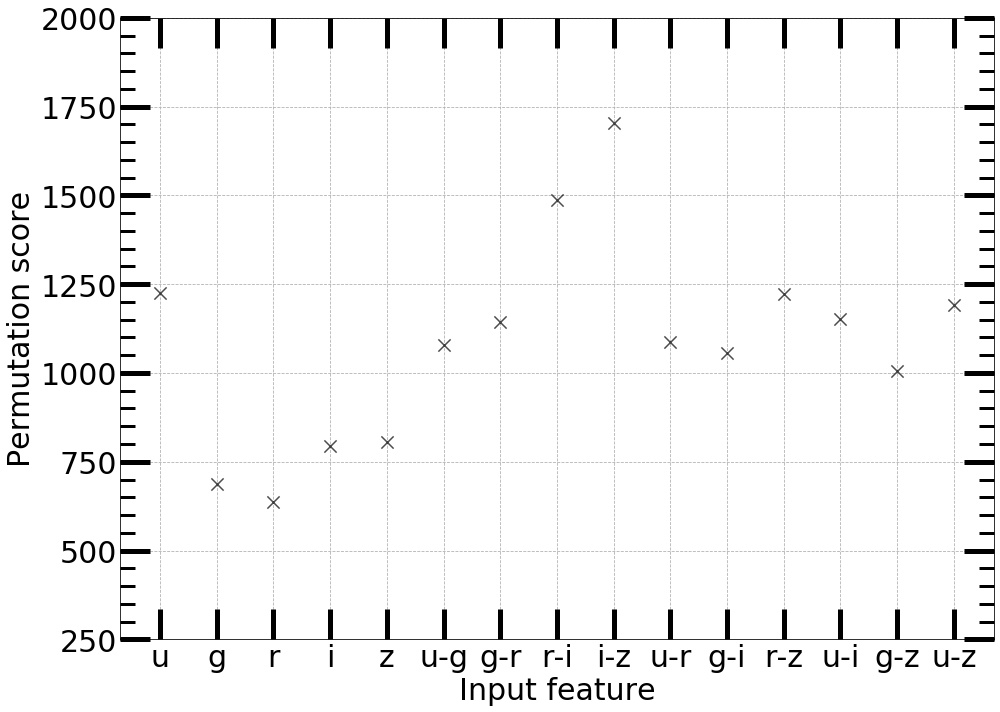}	
	\caption{This figure shows the importance (N.B. a lower permutation score signifies greater importance) of each input feature to our background subtraction model, where the permutation score was based on the number of `actual' cluster galaxies identified by our background subtraction model after randomly shuffling the data for each input feature.}
    \label{fig:test_feature_importance}
\end{figure}

In Figure \ref{fig:test_feature_importance}, we examined the importance of each input feature to our background subtraction model. This involved randomly shuffling the data of each input feature and then applying our background subtraction model on the dataset to observe how the shuffled feature impacted the predictive performance. This strategy is known as permutation feature importance testing \citep{random_forest}, where the permutation scores were based on the number of `actual' cluster galaxies identified by our background subtraction model. In particular, a lower permutation score for an input feature implied greater reliance of our background subtraction model on that specific input feature to provide good predictive performance, because randomly shuffling the data for an important input feature would result in fewer `actual' cluster galaxies being identified. We applied this permutation feature importance test to galaxies in our test set, which originally contained $3750$ cluster galaxies. Subsequently, we observed that $g$, $r$, $i$, $z$, $u-g$, $u-r$, $g-i$ and $g-z$ appeared to have greater significance to our background subtraction model whereas $u$, $g-r$, $r-i$, $i-z$, $r-z$, $u-i$ and $u-z$ appeared to have lesser significance to our background subtraction model. Although, it is important to note that our background subtraction model had effectively utilised all the input features since the number of identified `actual' cluster galaxies for each input feature was still only a fraction of the original number of `actual' cluster galaxies.

\section{Discussion}
\label{sec:discussion}

In Figure \ref{fig:photometric_cluster_and_field_galaxies_histogram}, we observed that the photometric redshift distribution of galaxies in our cluster galaxy sample was skewed towards higher redshifts, such that higher redshift cluster galaxies were overrepresented. To achieve a fair representation of cluster galaxies at different redshifts in our background subtraction model, we randomly sampled a fixed number of cluster galaxies within fixed redshift bin sizes of $0.01$ when creating our training, validation and test sets. This ensured that our background subtraction model was exposed to equal numbers of cluster galaxies at various redshifts within colour-magnitude space. We also exposed our background subtraction model to equal numbers of cluster and field galaxies during its training. This was to ensure a fair representation of the different galaxy classes in our background subtraction model.

We initially constrained our training sample to only spectroscopically confirmed cluster galaxies from the AMF11 catalogue but we quickly noticed that the training sample itself had a significant drop in the number of faint cluster galaxies across all redshifts when compared to the non-spectroscopically confirmed cluster galaxies. As such, we decided not to adopt this constraint when training our background subtraction model. Furthermore, we did not utilise spectroscopically confirmed field galaxies since it was difficult to acquire a sample that was representative of all potential foreground and background galaxies encountered within a random field. Although, in future work this may be possible since the number of spectroscopically confirmed cluster and field galaxies would naturally increase over time.

When constructing our scaling relation, we employed `actual' $r_{200}$ values that were estimated from a scaling relation (see Equation $1$ in \cite{whl12}) that was based on the total $r$ filter luminosity of identified cluster galaxies within a $2.5$ Mpc radius from the cluster center at each clusters redshift. This meant that the errors from their estimated $r_{200}$ values would have carried over into our estimated $r_{200}$ values too. In future work, we could instead consider employing $r_{200}$ values from X-ray cluster catalogues as X-ray emission measurements are not as significantly influenced by projection effects \citep{xray_vs_optical}. This would improve the overall precision of our `actual' $r_{200}$ values and thus improve the precision of our cluster richness estimates within $r_{200}$. Furthermore, we can establish a scaling relation for any radii, not only $r_{200}$, as long as we have sufficient data to enable the construction of a scaling relation for the radii.

In Figures \ref{fig:train_optimal_pred_richness} and \ref{fig:test_optimal_pred_richness} we did not observe any redshift biases in our estimated cluster richnesses after we applied completeness corrections to account for fewer observed galaxies at the faint end of the luminosity distribution of individual clusters. This indicated that incompleteness of our cluster galaxy sample from the AMF11 catalogue had a bigger impact on estimating cluster richnesses than incompleteness from misclassifications by our background subtraction model. Although, in Figures \ref{fig:train_scaling_r200}, \ref{fig:ae_recovered_redshift_test} and \ref{fig:test_scaling_r200} we observed a larger drop in the number of identified cluster galaxies at higher redshifts (i.e. $z > 0.3$) when compared to the number of identified cluster galaxies at lower redshifts. We believed that this could be due to the cluster galaxies at higher redshifts having larger photometric errors than cluster galaxies at lower redshifts, which can be seen in Figure \ref{fig:ae_recovered_colour_test} by the increased scatter between data points as redshift increased. Naturally, this would make it more difficult for our background subtraction model to identify them. As such, we would expect there to be fewer cluster galaxies identified at higher redshifts, since we did not truly account for cluster galaxies having larger photometric errors at higher redshifts in our background subtraction model. In the future, it would be beneficial to obtain and utilise a larger cluster galaxy sample when training our background subtraction model, which would hopefully reduce this effect by exposing the model to more examples. It may also be beneficial to employ an algorithm that can learn to interpolate regions within colour-magnitude space in order to account for the larger photometric errors at higher redshifts, such as a variational autoencoder \citep{variational_autoencoder}.

In this work, we used the Schechter function to fit to the luminosity distribution of identified cluster galaxies, where it was important to review the cluster membership status of each individual galaxy in order to minimise severe contamination from bright interloping field galaxies when fitting the Schechter function. Although, we were aware of alternative luminosity functions that could be used to fit to the luminosity distribution of cluster galaxies. Two other commonly used luminosity functions\footnote[19]{We recommend the reader to refer to \cite{richness_independent_of_redshift} for an overview of different luminosity functions.} include the Zwicky function \citep{zwicky_function} and Abell function \citep{abell_function}. Briefly, the Zwicky function is fitted by considering the difference in magnitude of each cluster galaxy from the brightest cluster galaxy whereas the Abell function is fitted by combining two separately fitted analytical functions. This means that the Zwicky function requires identifying the brightest cluster galaxy beforehand whereas the Abell function is not continuous at all luminosities. We decided to use the Schechter function over these other luminosity functions because the Schechter function did not have strict prerequisite conditions and offers continuity (i.e. it was composed of a power law and an exponential function) at all luminosities \citep{richness_independent_of_redshift}.

We fitted the Schechter function to a composite luminosity distribution that consisted of a subsample of identified cluster galaxies with high completeness to obtain best fit parameter values of $M^{*}$ = $-22.81$ with a standard deviation of $\pm 0.5$; $n^{*}$ = $159.82$ with a standard deviation of $\pm 154.62$ and $\alpha$ = $-1.99$ with a standard deviation of $\pm 0.37$. We did not allow $\alpha$ to be greater than $-1$ or lesser than $-2$ when performing Chi-squared fitting, as we assumed that it would be unphysical for the number of cluster galaxies to be decreasing or increasing rapidly at fainter magnitudes respectively. We also did not set any specific bounds for $M^{*}$ and $n^{*}$, since these parameters were more dependent on the given data. We attempted to compare our best fit parameter values for $M^{*}$ and $\alpha$ to the best fit parameter values of $M^{*}$ and $\alpha$ found in the literature from cluster studies to determine whether our best fit parameter values for $M^{*}$ and $\alpha$ were appropriate as `universal' values. However, we found that the literature contained a wide range of values for $M^{*}$ and $\alpha$ that depended on a variety of different factors (e.g. photometric system used, magnitude range examined, redshift range examined, cluster mass range examined, composition of galaxy types in cluster sample, background subtraction method used). Although, we noticed that some typical values obtained for $M^{*}$ and $\alpha$ span approximately from $-23$ to $-20$ and $-2.1$ to $-0.8$ respectively (e.g. \citealt{schechter_alpha_0}; \citealt{schechter_alpha_1}; \citealt{schechter_alpha_2}; \citealt{schechter_alpha_3}; \citealt{schechter_alpha_4}; \citealt{schechter_alpha_5}). This suggested that our assumptions for $M^{*}$ and $\alpha$ were not unreasonable.

We remind the reader that our approach for estimating cluster richness was based only on the number of cluster galaxies identified by our background subtraction model within a defined magnitude range and given search area. This would be particularly beneficial for cosmological studies \citep{richness_independent_of_redshift}, such as comparing simulated and observed halo mass functions (e.g. \citealt{comparing_halo_mass_function_0}; \citealt{comparing_halo_mass_function_1}), since it would reduce the complexity of modeling an appropriate selection function to correct for biases from post-processing (e.g. incorrect star/galaxy classification, deblending/interpolation issues, misestimated photometric redshifts) or survey conditions (e.g. flux limitations, oversaturation by bright stars, different aperture sizes) \citep{selection_function}. In addition, our background subtraction method provides robustness when estimating cluster richness along any line-of-sight environment since it assesses the cluster membership status of galaxies based only on their photometric measurements. This is not easily achievable when using simple statistical-based or colour-based background subtraction methods. Furthermore, our background subtraction method does not require us to make any assumptions about the properties of the cluster and field galaxies since these properties are self-learned by the AE algorithm. This means that our background subtraction method is not intrinsically biased towards selecting different galaxy types.

In future work, it would be interesting to examine the applicability of our background subtraction method on different usage cases. These include studying the properties and evolution of identified cluster galaxies or deciding spectroscopic follow-ups of potential galaxy members in clusters or measuring the observed radial density, luminosity and redshift profiles of clusters. We also aim to extend this current work by also establishing an empirical scaling relation between our richness estimates and cluster dark matter halo masses, that have been inferred via weak gravitational lensing, in order to construct an observed halo mass function for constraining cosmological parameters. In addition, we intend to integrate our background subtraction method with our own cluster finder model (Deep-CEE, \citealt{deep-cee}) and photometric redshift estimator model (Z-Sequence, \citealt{z-sequence}) to mask or remove interloping line-of-sight galaxies in image data or photometric catalogue data respectively to further minimise their model predictions errors.

We note that there are various other types of conventional machine learning algorithms\footnote[20]{We recommend the reader to refer to \url{https://pyod.readthedocs.io/en/latest/pyod.models.html} for an extensive list of outlier detection algorithms \citep{pyod}.} available, aside from AE's, which could be used for the task of performing background subtraction. These could include the K-nearest neighbours algorithm \citep{knn}, K-means algorithm \citep{kmeans}, isolation forest algorithm \citep{isolation_forest}, support vector machine algorithm \citep{svm} and XGBoost algorithm \citep{xgboost}. The reason we chose to utilise an AE over other conventional machine learning algorithms was due to the fact that an AE is a deep neural network, which is capable of self-learning the importance of input features. On the other hand, most conventional machine learning algorithms require important features to be manually extracted in order to attain good predictive performance, which can be time-consuming and difficult to do when there are many complex features (\citealt{machine_learning_vs_deep_learning_0}; \citealt{machine_learning_vs_deep_learning_1}; \citealt{machine_learning_vs_deep_learning_2}; \citealt{machine_learning_vs_deep_learning_3}). 

When training our background subtraction model, we used a Monte Carlo cross-validation strategy to determine an optimal hyper-parameter combination that offered the best predictive performance possible across different weight initialisations by performing random subsampling of our training and validation sets. Although, this may have resulted in some galaxies not being utilised at all (i.e. if the galaxy was not randomly chosen to be in any of our training, validation or test sets), which is not maximising data efficiency. In future work, we could employ a k-fold cross-validation strategy \citep{k-fold_cv} for hyper-parameter tuning and model evaluation. This would improve data efficiency and model generalisation as our background subtraction model would be evenly examined across all available data during its training and testing phases.

We performed permutation feature importance testing to determine which input features were deemed as important by our background subtraction model when identifying `actual' cluster galaxies. From which, we found that the following input features displayed high significance: $g$, $r$, $i$, $z$, $u-g$, $u-r$, $g-i$ and $g-z$. This tells us that our background subtraction model had learned to utilise most of the available photometric information in high dimensional colour-magnitude space. This was more efficient than only utilising a two dimensional colour-magnitude diagram, which is typically used when attempting to detect cluster galaxies within colour-magnitude space (e.g. \citealt{red_sequence_properties_0}; \citealt{red_sequence_properties_1}; \citealt{red_sequence_properties_2}; \citealt{red_sequence_properties_3}). We believe that these specific input features were important to our background subtraction model due to two main reasons. Firstly, in Figure \ref{fig:cmd_actual_vs_predict_test} it can be seen that the majority of the cluster and field galaxy population can be distinguished via filter magnitudes within colour-magnitude space. This explained why our background subtraction model prioritised several filter magnitudes when performing background subtraction. Secondly, in Figure \ref{fig:cmd_actual_vs_predict_test} it can also be seen that a minority of cluster galaxies overlapped with field galaxies within colour magnitude space. This explained why our background subtraction model also prioritised several colours in combination with the filter magnitudes to distinguish between these overlapping cluster and field galaxies. Based on these reasons, it is not unreasonable to assume that our background subtraction model can recognise the broad spectral features (e.g. $4000\AA$ break) and overall shape of the observed spectral energy distribution\footnote[21]{We recommend the reader to refer to \cite{galaxy_spectra} for further details on the observed spectral energy distribution of different galaxies.} of cluster galaxies\footnote[22]{Clusters typically have a majority population of elliptical and lenticular galaxies with a minority population of spiral galaxies \citep{galaxy_types}.} at different redshifts.

In future work, it would be interesting to examine the impact from including additional features such as galaxy sizes, morphology and surface brightness as inputs for our background subtraction model. However, we note that we cannot easily reapply our method to galaxy surveys that do not readily provide information for all our required input features. Furthermore, our background subtraction model is not provided with redshift information as an input feature when distinguishing between cluster and field galaxies. Instead, we wanted our background subtraction model to self-learn about the photometric properties of cluster galaxies belonging to different redshift intervals, which is similar to how photometric redshifts of individual galaxies are estimated by empirical algorithms.

\section{Conclusion}
\label{sec:conclusion}

We present a proof-of-concept study of AutoEnRichness, a hybrid empirical and analytical approach that uses a multi-stage machine learning algorithm and a conventional luminosity distribution fitting approach to perform background subtraction and estimate cluster richnesses respectively. We utilised photometric data from the SDSS-IV DR16 to train our background subtraction model, which learned to reconstruct the photometry of cluster galaxies in order to distinguish between cluster and field galaxies. We then examined the predictive performance of our background subtraction model at distinguishing between cluster and field galaxies in a test set, which resulted in a balanced accuracy of $83.20$ per cent. Subsequently, we constructed a scaling relation that estimated $r_{200}$ when given the number of cluster galaxies identified by our background subtraction model within a search radius of $2.5$ Mpc at each cluster's spectroscopic redshift. We utilised this learned scaling relation to resample galaxies within an $r_{200}$ radius for each cluster. Next, we fitted the Schechter function to a composite luminosity distribution that consisted of a subsample of cluster galaxies identified by our background subtraction model within $r_{200}$ that had high completeness. We used a Chi-squared fitting approach to determine an optimal $r$ filter absolute magnitude bin size of $0.52$ and best fit parameter values of $M^{*}$ = $-22.81$ with a standard deviation of $\pm 0.5$; $n^{*}$ = $159.82$ with a standard deviation of $\pm 154.62$ and $\alpha$ = $-1.99$ with a standard deviation of $\pm 0.37$. We then used the optimal $r$ filter absolute magnitude bin size and best fit parameter values for $M^{*}$ and $\alpha$ to fit the Schechter function to the luminosity distribution of individual clusters. We estimated cluster richnesses within $r_{200}$ by computing the integral of the locally fit Schechter function. Lastly, we applied the optimal $r$ filter absolute magnitude bin size and best fit parameter values for $M^{*}$ and $\alpha$ to another test set of clusters to obtain a median absolute percentage error of $33.50$ per cent between our estimated cluster richnesses and WH15 richnesses within $r_{200}$. We note that the only cluster prerequisites for AutoEnRichness were the astronomical coordinates of the approximate cluster location as well as an initial cluster redshift estimate for computing appropriate cluster radii. We intend for AutoEnRichness to be combined with the Deep-CEE \citep{deep-cee} and Z-Sequence \citep{z-sequence} algorithms to obtain the key measurements (i.e. position from cluster detection and distance from redshift estimation respectively) needed for conducting astrophysics and cosmology research. In future work, it would be beneficial to develop a data pipeline that integrates AutoEnRichness with these other methods into an end-to-end process in preparation for usage on upcoming large-scale galaxy surveys, such as the Legacy Survey of Space and Time \citep{lsst_survey} and $Euclid$ (\citealt{euclid_survey_0}; \citealt{euclid_survey_1}).

\section*{Acknowledgements}

We would like to thank the anonymous referee for their thorough feedback which has improved the clarity of our paper.

We gratefully acknowledge the support from the Science and Technologies Facilities Council studentship funding. We thank members of the XMM Cluster Survey for useful discussions. We would also like to thank the developers of \textsc{Vizier} \citep{vizier}, \textsc{TOPCAT} \citep{TOPCAT}, James Schombert at the University of Oregon, Edward L. Wright at the University of California, Los Angeles \citep{cosmo_calc}, \textsc{Scikit-Learn} \citep{scikit-learn}, \textsc{SciPy} \citep{scipy} and \textsc{TensorFlow} \citep{tensorflow} for allowing the open distribution and free usage of their software for research.

Funding for the Sloan Digital Sky Survey IV has been provided by the Alfred P. Sloan Foundation, the U.S. Department of Energy Office of Science, and the Participating Institutions. 

SDSS-IV acknowledges support and resources from the Center for High Performance Computing at the University of Utah. The SDSS website is www.sdss.org.

SDSS-IV is managed by the Astrophysical Research Consortium for the Participating Institutions of the SDSS Collaboration including the Brazilian Participation Group, the Carnegie Institution for Science, Carnegie Mellon University, Center for Astrophysics | Harvard \& Smithsonian, the Chilean Participation Group, the French Participation Group, Instituto de Astrof\'isica de Canarias, The Johns Hopkins University, Kavli Institute for the Physics and Mathematics of the Universe (IPMU) / University of Tokyo, the Korean Participation Group, Lawrence Berkeley National Laboratory, Leibniz Institut f\"ur Astrophysik Potsdam (AIP), Max-Planck-Institut f\"ur Astronomie (MPIA Heidelberg), Max-Planck-Institut f\"ur Astrophysik (MPA Garching), Max-Planck-Institut f\"ur Extraterrestrische Physik (MPE), National Astronomical Observatories of China, New Mexico State University, New York University, University of Notre Dame, Observat\'ario Nacional / MCTI, The Ohio State University, Pennsylvania State University, Shanghai Astronomical Observatory, United Kingdom Participation Group, Universidad Nacional Aut\'onoma de M\'exico, University of Arizona, University of Colorado Boulder, University of Oxford, University of Portsmouth, University of Utah, University of Virginia, University of Washington, University of Wisconsin, Vanderbilt University, and Yale University.

\section*{Data Availability}

The AMF11 catalogue \citep{photometric_cluster_members_0} of individual cluster galaxies is publicly available on Vizier via: \url{http://vizier.u-strasbg.fr/viz-bin/VizieR-3?-source=J/ApJ/736/21/mg}. The SDSS-IV DR16  \citep{sdss_dr16} photometry data of individual galaxies is also publicly available on Vizier via: \url{http://vizier.u-strasbg.fr/viz-bin/VizieR?-source=V/154}. Furthermore, the WH15 \citep{wh15} and redMaPPer v6.3 \citep{redmapper} cluster catalogues can be found publicly via: \url{http://vizier.u-strasbg.fr/viz-bin/VizieR-3?-source=J/ApJ/807/178/table3} and \url{http://risa.stanford.edu/redmapper/}.

%%%%%%%%%%%%%%%%%%%%%%%%%%%%%%%%%%%%%%%%%%%%%%%%%%


%\bibliographystyle{mnras}
%\bibliography{references} % if your bibtex file is called example.bib

%%%%%%%%%%%%%%%%%%%%%%%%%%%%%%%%%%%%%%%%%%%%%%%%%%

%%%%%%%%%%%%%%%%% APPENDICES %%%%%%%%%%%%%%%%%%%%%

%\appendix

%\section{Appendix}
%\label{sec:Appendix}

%%%%%%%%%%%%%%%%%%%%%%%%%%%%%%%%%%%%%%%%%%%%%%%%%%

% Don't change these lines
\bsp	% typesetting comment
\label{lastpage}
\end{document}